\documentclass[journal]{IEEEtran}

\usepackage{cite}

\ifCLASSINFOpdf
\else
\fi
 \usepackage[dvips]{graphicx}
\usepackage{color}
\usepackage{subfigure}
\usepackage{url}
\usepackage[cmex10]{amsmath}
\usepackage{amssymb}
\newtheorem{Theorem}{Theorem}
\newtheorem{Lemma}{Lemma}

\usepackage{amsmath}
\DeclareMathOperator*{\argmax}{arg\,max}
\usepackage{algorithm,algorithmic}
\allowdisplaybreaks

\begin{document}
\definecolor{MyDarkBlue}{rgb}{0,0.08,0.45}
\title{Cross-Layer Scheduling for OFDMA-based Cognitive Radio Systems with Delay and Security Constraints}
%
%
%

\author{Xingzheng~Zhu, Bo~Yang, Cailian~Chen, Liang~Xue, Xinping~Guan, Fan~Wu
\thanks{Copyright (c) 2013 IEEE. Personal use of this material is permitted. However, permission to use this material for any other purposes must be obtained from the IEEE by sending a request to pubs-permissions@ieee.org.}
\thanks{X. Zhu, B. Yang, C. Chen and X. Guan are with Department of Automation, Shanghai Jiao Tong University, and Key Laboratory of System Control and Information Processing, Ministry of Education of China, Shanghai 200240, China  (Emails:\{wendyzhu, bo.yang, cailianchen, xpguan\}@sjtu.edu.cn). B. Yang, C. Chen and X. Guan are also with the Cyber Joint Innovation Center, Hangzhou, China.  L. Xue is with School of Information and Electrical Engineering, Hebei University of Engineering, Handan 056038, China (Email: liangxue@hebeu.edu.cn). F. Wu is with the Department of Computer Science
and Engineering, Shanghai Key Laboratory of Scalable Computing
and Systems, Shanghai Jiao Tong University, Shanghai 200240, China (E-mail: fwu@cs.sjtu.edu.cn). B. Yang is the corresponding author of this paper.}
}

\maketitle
{ 
\begin{abstract}
 This paper considers the resource allocation problem in an Orthogonal Frequency Division Multiple Access (OFDMA) based cognitive radio (CR) network, where the CR base station adopts full overlay scheme to transmit both private and open information to multiple users with average delay and power constraints. A stochastic optimization problem is formulated to develop flow control and radio resource allocation in order to maximize the long-term system throughput of open and private information in CR system and ensure the stability of primary system. The corresponding optimal condition for employing full overlay is derived in the context of concurrent transmission of open and private information. An online resource allocation scheme is designed to adapt the transmission of open and private information based on monitoring the status of primary system as well as the channel and queue states in the CR network. The scheme is proven to be asymptotically optimal in solving the stochastic optimization problem without knowing any statistical information. Simulations are provided to verify the analytical results and efficiency of the scheme.
\end{abstract}}


%
\IEEEpeerreviewmaketitle

\section{Introduction}

%
%
%
%
\IEEEPARstart{T}{he}  emergency of high-speed wireless applications and  increasing scarcity of available spectrum remind
researchers of spectrum utilizing efficiency. The concept of CR  provides the potential
technology in increasing spectrum utilizing efficiency\cite{weiss2004spectrum,huang2011coolest}  because CR allows unlicensed users (also known as secondary users (SUs)) to access some
spectrum which is already allocated to primary user (PU) or licensed user who has the authority to access the spectrum by spectrum sensing \cite{6464545, 6099643}. As
another promising technology of high speed wireless communication system,  OFDMA   is a candidate for CR systems  \cite{weiss2004spectrum} due to its flexibility in allocating spectrum among SUs
\cite{lawrey1999multiuser}. Hence, OFDMA-based CR networks are catching great attention
\cite{zhou2011multiuser,zhou2011low}.  This paper focuses on an OFDMA-based CR network without loss of generality.

In order to exploit the capacity of the whole OFDMA-bsed CR system, this paper aims at maximizing the secondary network
capacity in consideration of the whole system transmission efficiency. Thus, the following three main issues should be considered.

Firstly, an efficient spectrum sharing scheme is  essential for exploiting the unused
spectrum  in OFDMA-based CR network. When a SU wants to access some spectrum,  it must ensure that the spectrum is
not accessed by any PU or  adapt its parameter to limit the  interference to PU. Both of these two mentioned spectrum utilization manners, known as overlay and underlay schemes, are conservative in some ways, since they ignore the PU's ability to tolerate some inference.

Secondly, due to that CR networks as well as many other kinds of wireless communication systems have a nature of broadcast,
security issues at physical layer have always been unavoidable in designing CR systems. Furthermore, to SUs, it is obviously practical that there exist both private and open transmission requirements. Then, the scheduling
among these two different kinds of transmission should be considered. In addtion,  delay performance is an indispensable quality of service (QoS) index in scheduling different transmissions.

Last but not least, the dynamic nature of OFDMA-based CR communication system brings another big challenge.  The random arrival of user
requests (from both PU and SU) and time-varying channel states renders dynamic resource allocation instead of fixed ones in exploiting the OFDMA secondary network capacity.

 Aiming at the above issues, the contributions of this paper are threehold:
 \begin{itemize}
\item First, this paper adopts a novel full overlay spectrum accessing scheme by exploiting PU's tolerance to interference.  Besides, the theoretic proof of  full { overlay's} optimality is
    given in the presence of both open and private transmissions.

\item Second, a joint encoding model is introduced to allow both private and open transmissions towards SUs with the full overlay spectrum sharing scheme. A dynamic resource allocation scheme consisting of flow control and radio resource allocation is developed by solving a formulated stochastic optimization problem under the delay and power constraints.

 \item Third, the proposed dynamic resource allocation scheme is proven to be close to optimality although its implementation only depending on instantaneous information.
\end{itemize}

This paper is organized as follows. Section II presents the related work. In Section III, we  introduce the system model and
relevant constraints in detail. Section IV formulates the problem. In Section V, we introduce our cross-layer optimization
algorithm.  We give the
performance bound and stability results in Section VI. Two different implementations are proposed in Section VII. In Section
VIII, some simulation results are shown. Finally, we conclude this paper in Section IX.

\section{Related Work}
There have been many works on spectrum sharing in OFDMA-based CR
networks\cite{almalfouh2011interference,zhang2009resource,zhang2009cross}. According to
\cite{wang2011advances,zhao2007survey}, the access technology of the SUs can be divided in two
categories: spectrum underlay and spectrum overlay. The first category means that SUs can access licensed spectrum during
PUs' transmission, while as is mentioned in \cite{zhao2007survey}, this approach imposes severe constraints on the
transmission power of SUs such that they can operate below the noise floor of PUs, e.g, in
\cite{huang2005spectrum,le2007qos,almalfouh2011interference}.  The second category means that SUs can only
access licensed spectrum when the PU is idle, e.g,
in\cite{levorato2012cognitive,huang2010distributed,georgiadis2006resource,zhang2009resource,zhang2009cross}. Considering
both these two strategies suffer from some drawbacks, the authors in  \cite{lapiccirella2013distributed} propose a new cognitive overlay
scheme requiring SUs to assess and control their interference impacts on PUs. In general, the cognitive base station (CBS) controls the
aggregate interference to primary transmission by allowing SUs to monitor channel quality indicators (CQIs), power-control
notifications and ACK/NAK of primary transmission. In this paper, this novel thought is extended into an OFDMA-based
CR system.

 On the other hand, dynamic resource allocation plays a critical  role in exploiting OFDMA network capacity.  The overall performance as well as the multiuser diversity of the system  can be improved by proper dynamic resource allocation
 \cite{jang2003transmit,kim2005downlink,seong2006optimal,zou2010network,xingzheng2012flow,shen2005adaptive,georgiadis2006resource,li2003dynamic}. Thus,
 dynamic resource allocation in OFDMA-based CR system has been attracting more attention recently. The corresponding spectrum sharing schemes in
 \cite{almalfouh2011interference,zhang2009resource,zhang2009cross} are all realized by dynamic resource
 allocation.

Besides the interference constraints, the works of delay aware transmission are also quite relative to this paper. Huang and Fang in \cite{huang2008multiconstrained} investigate both reliability and delay constraints in
 routing design for wireless sensor network.
Cui et al. in \cite{cui2012survey} summarize three approaches to deal with delay-aware resource allocation in wireless networks. A constrained predictive control strategy is
proposed in \cite{5572437} to compensate for network-induced delays with stability guarantee. Those three
methods are based on large deviation theory, Markov decision theory and Lyapunov optimization techniques.
As to the first two methods, they have to know some statistical  information on channel state and random arrival data rate to
design algorithm, while these prior knowledge is expensive to get, even unavailable. To overcome this problem, many authors
pay attention to Lyapunov optimization techniques. References \cite{xue2010delay} and \cite{urgaonkar2009delay} investigate scheduling in
multi-hop wireless networks and resource allocation in cooperative communications, respectively  as two typical applications
of Lyapunov optimization  in delay-limited system. In this paper, we utilize this tool to
dispose the resource allocation problem in OFDMA-based  CR networks.

As for secure transmission, Shannon's information theory  laid the foundation for information-theoretic
security\cite{shannon1949communication} and the concept of wire-tap channel  was proposed in \cite{ozarow1985wire}.  There has been
some research on exploiting  security capacity  in OFDMA network by dynamic resource allocation, such as in \cite{ng2012energy} and
\cite{wang2011power}. In CR area, the study of secure transmission from
information-theoretic aspect is very limited. Pei et al. in \cite{pei2010secure} first investigate  secrecy capacity of the secure
multiple-input single-output (MISO) CR channel. Kwon et al. in \cite{kwon2012secure} utilize the concept of security capacity to
explore MISO CR systems where the secondary system secures the primary communication in return for permission to use the
spectrum. Both these two works focus on only private message transmission. The security and common capacity of cognitive
interference channels is analyzed in \cite{liang2009capacity}. The entire capacity of a
MIMO broadcast channel with common and confidential messages is obtained in \cite{ekrem2012capacity}. The paper \cite{conf/wcnc/ZhuYG13} considers the problem of optimizing the
security and common capacity of  an OFDMA downlink system by dynamic resource allocation.  This paper further considers the transmissions of private and open flows in CR networks with delay constraints.

\section{System Model}
\begin{figure}
  \centering
  \includegraphics[width=3in, angle=270]{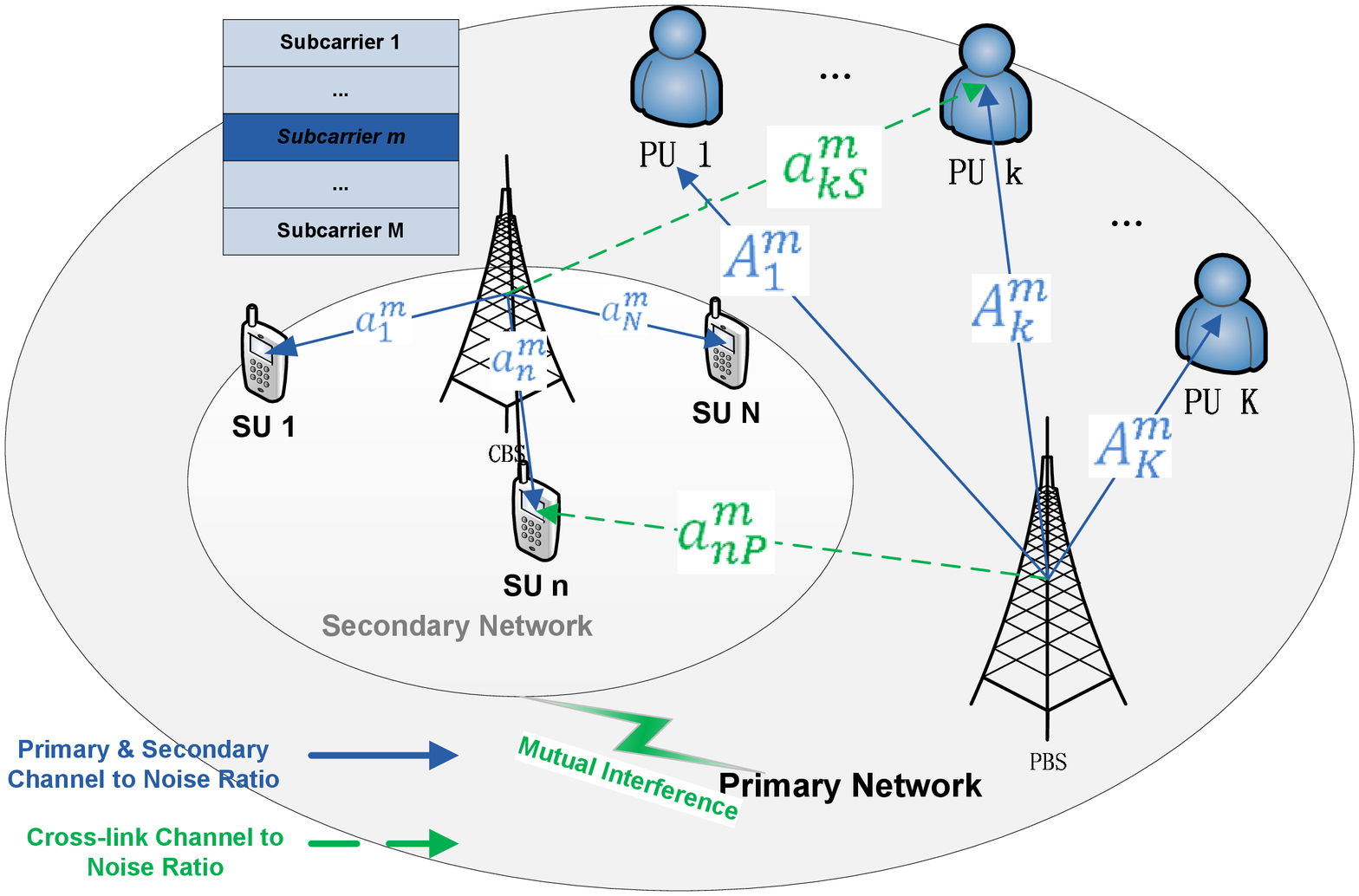}\\
  \caption{ General network model}\label{sysfig:2}
\end{figure}

{ The system model consists of  multiple primary  links and multiple secondary links as Fig. \ref{sysfig:2} shows. The total
bandwidth $B$ is divided into $M$ subcarriers equally using Orthogonal Frequency Division Multiplexing (OFDM). Assume that $M=B$ holds for simplicity of expression. The subcarrier set of the network is denoted as $\mathbf{M}=\{1,2,\cdots,M\}$ and $m\in \mathbf{M}$ denotes subcarrier index. The downlink case is considered. The primary
link is from a single primary base station (PBS) to $K$ PUs. Secondary links are from a common CBS to $N$ SUs. We denote $k\in\{1,2,\cdots,K\}$ and $n\in\{1,2,\cdots,N\}$ as the indexes of PU and SU respectively.
The system operates in slotted time, and $T$ is the length of a time slot. Hereafter, $[tT,(t+1)T)$ is just denoted by $t$ for  brevity.}

The set of subcarriers occupied by PU $k$ on timeslot $t$ is denoted as
$\mathbf{\Gamma}_k^{PU}(t)=\{\tau_1^k(t),\tau_2^k(t),\cdots,\tau_{m^k(t)}^k(t)\}$ where $m^k(t)$ is the number of subcarriers occupied
by PU $k$ and $\mathbf{\Gamma}_k^{PU}(t)\subseteq\{1,2,\cdots,M\}$. The power
set $\mathbf{P}_k^{PU}(t)=\{P_k^m(t)|m\in\mathbf{M}\}$ is the set of transmission power from PBS to PU $k$, where for $m\in\mathbf{\Gamma}_k^{PU}(t)$, $ P_k^m(t)>0$, else $P_k^m(t)=0$. For brevity, we will omit the time index $(t)$ somewhere in further discussion.
$\boldsymbol{P}^{SU}=\{p_n^m|\forall n,\forall m\}$ denotes the overall SUs power allocation policy set and $p_n^m$
 represents the power allocated by CBS to user $n$ in subcarrier $m$. Denote ${\mathbf{\Gamma}}_n^{SU}=\{\varpi_n^m|\forall m\}$ as the subcarrier
assignment
policy of SU $n$, where $\varpi_{n}^{m}$ is either 1 representing subcarrier $m$ is assigned to SU $n$, or 0 otherwise.
Then let ${\mathbf{\Gamma}^{SU}}=\{{\mathbf{\Gamma}}_n^{SU}(t),\forall n\}$ be the overall subcarrier assignment policy of secondary network.

{Due to the orthogonal properties of OFDMA
technology, there exists no mutual influence between every two SUs. However, there exists mutual interference between the primary and secondary networks when PU and SU access in the same subcarrier.}

The channel gains include the one of secondary user $n$ on subcarrier $m$, $h_n^m$ and the one of
primary user $k$ on subcarrier $m$, $H_k^m$. The additive white gaussian noise (AWGN) is $\sigma^2$. The
corresponding subcarrier gain-to-noise-ratio\footnote[1]{Also called gain-to-noise-plus-interference-ratio when SU and PU access in the same subcarrier.} (C/I) in slot $t$ are thus
defined as $a_n^m(t)=\frac{{h_n^m(t)}^2}{\sigma^2}$ and $A_k^m=\frac{{H_k^m(t)}^2}{\sigma^2}$ respectively as  illustrated in Fig.\ref{sysfig:2}.  The set
$\boldsymbol{a}(t)=\{A_k^m(t),a_{n}^{m}(t),\forall n, \forall
 m, \forall k\}$ represents the system channel state information (CSI). All channels are assumed to be slow fading, and thus
 $\boldsymbol{a}(t)$ remains fixed during one slot and changes between two \cite{tse2005fundamentals}. { In this work, there exists an reasonable assumption that the system CSI is known to BS. As in \cite{wallace2002method}, BS can get  full-CSI by utilizing pilot symbols and CSI feedback process.  Besides,  at the beginning of every slot, PU reports
 $P_k^mA_k^m$ to PBS. For example, the PU reports a received-signal-strength index to PBS in packets such as RSSI
 reports. We assume the CBS will listen to the information to derive $P_k^mA_k^m$ before accessing subcarrier $m$ { \cite{lapiccirella2013distributed,wallace2002method}}.

  Denote ${h_{kS}^m}$ as  the cross-link interference channel gain from CBS to PU $k$ on subcarrier $m$ and let $a_{kS}^m=\frac{{h_{kS}^m}^2}{\sigma^2}$. Similarly, denote ${h_{nP}^m}$ as  the cross-link interference channel gain from PBS to SU $n$ on subcarrier $m$ and let $a_{nP}^m=\frac{{h_{nP}^m}^2}{\sigma^2}$.  It is assumed that $a_{kS}^m$ and $a_{nP}^m$ can be got by the CBS. $a_{kS}^m$ can be estimated
 by CBS from the PU feedback signal based on reciprocity. $a_{nP}^m$ can be estimated by SUs through training
 and sensing and the estimation results are sent to CBS \cite{wallace2002method}.  Beyond that, information about cross-link channel state could also be measured periodically by a band manager either \cite{suraweera2010capacity,almalfouh2011interference}.}

\begin{figure}
  \centering
  \includegraphics[width=2.5in, angle=270]{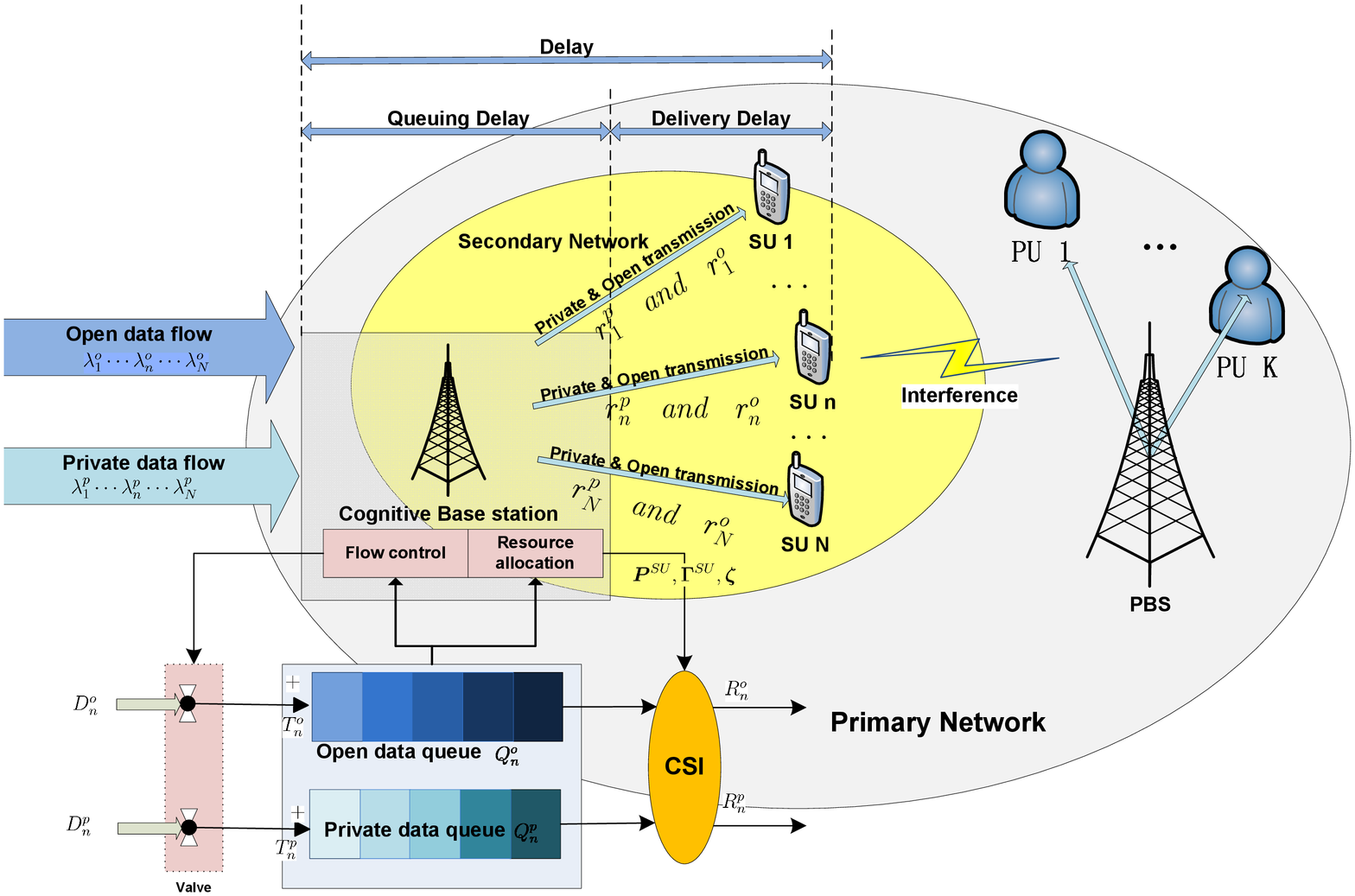}\\
  \caption{ Transmission model of secondary network}\label{sysfig:3}
\end{figure}

Compared to pervious work, this paper considers a more complicated and  practical situation of SU transmission.  The CBS
transmits both private and open data to each SU as Fig.\ref{sysfig:3} shows. The private data has security requirement and open data has long-term time-average delay constraint. Instead of that both open and private data have delay constraint, only delay constraints on open transmissions are considered in this paper for simplifying the mathematic expressions, since the handling of delay constraint in secure transmission is totally the same as open transmission. Actually, in real wireless communication systems, there exists some private transmission having no strict delay constraint, e,g. updating  contact information in mobile devices. At the beginning of every time slot, random data packets arrive at CBS. CBS  decides whether to admit it into the system or not. Besides, CBS is also in charge of resource allocation to assign  power and subcarriers among SUs. CBS utilizes the information of data queue and CSI to allocate resources. The system performance can be optimized and the queuing delay of open data can be ensured to fulfill by flow control and resource allocation.

In the side of CBS, the amount of open data packet of SU $n$, $D_n^o(t)$, and private data, $D_n^p(t)$  that arrive at CBS
during slot $t$ are independent identically distributed (i.i.d) stochastic processes, e,g. Bernoulli processes, with the long-term average arrival rates $\lambda_n^o$ and $\lambda_n^p $, and their upper bounds are $\mu_{max}$ and $D_{max}$,
respectively.  These packets can not be transmitted to target users instantaneously due to the time-varying channel conditions and they are enqueued at the CBS. However, only parts of these packets are admitted into each queue towards each user for stability reason to be specified later. The  amounts of  open and private data admitted by respect queues are  $T_n^o(t)$ and $T_n^p(t)$ and  CBS is in charge of determining $T_n^o(t)$ and $T_n^p(t)$ according to a
certain principle which would be specified in Section V.

\subsection{Capacity model}
In OFDMA-based CR networks, SU and PU can access in the same subcarrier with mutual interference. However, due to the characteristic of OFDMA networks, each subcarrier can not be assigned  to more than solitary user in any secondary or primary network. Thus the following formulation is set to
ensure the limitation in CBS:
\begin{align}\label{assign}
0\le\sum_{n=1}^N\varpi_n^m\le1,\qquad\forall m
\end{align}

 CBS will realize the occupied subcarrier set $\boldsymbol{\Gamma}_k^{PU}=\{m|P_k^m>0,\forall m\}$, and we denote
 $\boldsymbol{\Gamma}_{SU}=\{1,2,\cdots,M\}-\bigcup_{k=1}^{K}\boldsymbol{\Gamma}_k^{PU}$. Thus the transmission rates of PU and SUs can be
 analysed by dividing $M$ subcarriers into two parts: one is $m\in \bigcup_{k=1}^{K}\boldsymbol{\Gamma}_k^{PU}$ where there exists interference
 between PU and SUs; another is $m\in\boldsymbol{\Gamma}_{SU}$ which means SUs can access these subcarriers without
 influencing primary link. Thus according to information theory the transmission rate of PU $k$ on subcarrier $m$ is:

 \begin{align}
 R_k^m=\left\{\begin{array}{ll}
 \log_2(1+\frac{P_k^mA_{k}^m}{1+a_{kS}^mp_{n'}^m})&m\in\boldsymbol{\Gamma}_k^{PU},n'\in\mathbf{\tilde{\Gamma}}_m\nonumber\\
 0&m\in\boldsymbol{\Gamma}_{SU}\nonumber
 \end{array}\right.
 \end{align}
 where  $\mathbf{\tilde{\Gamma}}_m$ is the set of SUs accessing
 subcarrier $m$. Furthermore, since in secondary network, only one SU can access one subcarrier, $n'$ is the only one
 element in set $\mathbf{\tilde{\Gamma}}_m$.

  It should be  noticed  that the total transmission rate in an OFDMA  network equals to the sum rates on all subcarriers.
 So the transmission rate of PU is:

 \begin{equation}
R_k^{PU}=\sum_{m\in\mathbf{\Gamma}_k^{PU}}R_k^m
 \end{equation}

 The channel capacities of SU $n$ on subcarrier $m$ can be expressed as:

 \begin{align}
C_n^m=\left\{\begin{array}{ll}
 \log_2(1+\frac{p_{n}^ma_{n}^m}{1+P_{k'}^ma_{nP}^m})&m\in\bigcup_{k=1}^K\boldsymbol{\Gamma}_k^{PU}, k'\in \mathbf{\hat{\Gamma}}_m\nonumber\\
 \log_2(1+p_{n}^ma_{n}^m)&m\in\boldsymbol{\Gamma}_{SU}\nonumber
 \end{array}\right.
 \end{align}
 where  $\mathbf{\hat{\Gamma}}_m$ is the set of PUs accessing
 subcarrier $m$. Furthermore, since only one PU can access one subcarrier, $k'$ is the only one
 element in set $\mathbf{\hat{\Gamma}}_m$. Denote $R_n^{SU}=\sum_mC_n^m$ as the sum transmission rate of SU $n$ without consideration of security.

 By introducing the joint transmission model, open and private data of one SU can be transmitted simultaneously.
{ Open message is jointly encoded with security message as random codes. In this way, although open message may be decoded by eavesdroppers, security message would be perfectly secure if the channel fading is properly utilized \cite{mclaughlin2014applications}. According to the theory of physical-layer security \cite{wang2011power}, if  the transmission rate of private data is less than \emph{security capacity}, the proposed joint-encoding model can at least realize physical-layer security in theory. \cite{mclaughlin2013secure,argon2013pre} propose physical-layer security realization applications using error correcting codes and pre-processor, which lays the foundation of realizing physical-layer security of the joint encoding model. }
 For each SU, CBS makes decision  if his
secure data could be transmitted in this slot
and this decision is expressed as the secure transmission control vector $\boldsymbol{\zeta}=(\zeta_{1},\zeta_{2},\cdots,\zeta_{N})$. The
indicator variable $\zeta_{n}=1$ implies that  private and open messages are encoded at rate $\hat{R}_{n}^p$ and
$R_n^{SU}-\hat{R}_n^p$  respectively in timeslot $t$  and $\zeta_{n}=0$ means that only open messages can be transmitted at
rate $R_n^{SU}$.

When CBS is transmitting private messages to SU $n$, all the other SUs except SU $n$ are treated as potential eavesdroppers { \cite{wang2011power}}.
According to
\cite{emre2011control}, subject to perfect private of SU $n$, the instantaneous private rate of SU $n$ on
subcarrier $m$ is the achievable channel capacity minus the highest eavesdropper capacity if there is no cooperation among
eavesdroppers. For each SU $n$, we define the
most potential eavesdropper on subcarrier $m$ as SU $\tilde{n}$ and $\tilde{n}=\argmax\limits_{n',n'\neq
n}a_{n'}^m$.  So the security capacity of SU $n$ on subcarrier $m$ is:
  \begin{equation}
  \hat{R}_n^{mp}=
  \begin{cases}
 [C_n^m-\log_2(1+\frac{p_{n}^mb_{n}^m}{1+P_{k'}^mb_{nP}^m})]^+& m\in\mathbf{\Gamma}_k^{PU}, k'\in \mathbf{\hat{\Gamma}}_m\\
  [C_n^m-\log_2(1+p_{n}^mb_{n}^m)]^+& m\in\mathbf{\Gamma}_{SU}\\
  \end{cases}
  \end{equation}
where $[\cdot]^+ =\max\{\cdot,0\}$, $b_n^m=a_{\tilde{n}}^{m}$ and $b_{nP}^m$ is the cross-link CSI
 from PBS to SU $\tilde{n}$ on subcarrier $m$. Obviously, $\hat{R}_n^p=\sum\limits_{m\in\mathbf{M}}\hat{R}_{n}^{mp}$.
Thus the achievable private rate of user $n$ is:
\begin{equation}
R_n^p=\zeta_n\hat{R}_{n}^p\nonumber
\end{equation}
and  the open rate of user $n$ is: $R_n^o=R_n^{SU}-R_n^p$.

\subsection{Queuing model}
There exist data queues in both PBS and CBS. Although we want to maximize the weighted throughput of SUs, PU queue
stability is a constraint in ensuring that PU's  long-term throughput is not affected by SU's transmission. It is assumed that the transmission rate of PBS
without interference is sufficient to serve PU's demand. However,  the primary network and  the secondary network will be influenced by each other if they work on the same channel. The transmission rate decrease of PU is due to the interference
brought by SU transmission, while the CBS can adjust its schedule to limit interference in order to ensure that PU's time-varying rate demands can be satisfied. Later, the notation of queue stability will be used to measure whether PU's demand can be fulfilled. In
\cite{lapiccirella2013distributed}, the interference is limited by that PU queue is kept stable under the influence caused
by the only one SU access. We continue to utilize this technique in scheduling our multi-SU access system.

First, it is necessary to introduce the concept of \emph{strong stability}. As a discrete time process,
$Q(t+1)=[Q(t)-S(t)]^++D(t)$ is \emph{strongly stable} if:
  \begin{equation}\label{stronglystable}
  \limsup_{t\rightarrow\infty}\frac{1}{t}\sum_{\tau=0}^{t-1}\mathbb{E}\{Q(\tau)\}<\infty
  \end{equation}
In particular, a  multi-queue  network is  stable when all queues of the network are \emph{strongly stable}.
According to \emph{Strong Stability Theorem} in \cite{neely2010stochastic}, for finite variable $S(t)$ and $D(t)$, strong
stability implies rate stability of $Q(t)$. The definition of rate stability can be found in \cite{neely2010stochastic}
and omitted here.

Furthermore, according to \emph{Rate Stability Theorem} in \cite{neely2010stochastic}, $Q(t)$ is rate stable if and only if $
d\le s $ holds where $d=\lim\limits_{t\rightarrow\infty}\frac{1}{t}\sum_{\tau=0}^{t-1}D(\tau)$ and
$s=\lim\limits_{t\rightarrow\infty}\frac{1}{t}\sum_{\tau=0}^{t-1}S(\tau)$.

Since the data can not be delivered instantly to PUs or SUs, there are data backlogs in the PBS and CBS waiting for transmitting to respective users.

\subsubsection{PU queue}
In PBS, the data queue of PU $k$ is updated as following:
  \begin{equation}
  Q_k(t+1)=[Q_k(t)-R_k^{PU}(t)]^++D_k^{PU}(t)
  \end{equation}
 where $D_k^{PU}(t)$ is the amount of  data packets randomly arriving at PBS during slot $t$ with the destination of PU $k$. We assume $D_k^{PU}(t)$ is
 an i.i.d stochastic process with its upper bound of $D_{max}^{PU}$ and its  long-term average arrival rates 
 $\lambda_k=\lim\limits_{t\rightarrow\infty}\frac{1}{t}\sum_{\tau=0}^{t-1}D_k^{PU}(\tau)$. As it has been mentioned before, $Q_k$ should be kept stable by limiting SUs' interference to
 primary link.  As  \emph{Rate Stability Theorem} shows,  $Q_k$ is rate stable if and only if $r_k^{PU}\ge\lambda_k$ where
 $r_k^{PU}\triangleq\lim\limits_{t\rightarrow\infty}\frac{1}{t}\sum_{\tau=0}^{t-1}R_k^{PU}(\tau)$. Therefore, if PU system is strongly stable, its long-term transmission is not affected by SUs.

 \subsubsection{SU data queues}
In CBS, there exist actual data queues of open and private data which are represented by $Q_n^o$ and $Q_n^p$ respectively for all
$n\in\{1,\cdots,N\}$. These queues  are updated as follows:
\begin{align}
 Q_n^o(t+1)&=[Q_n^o(t)-R_n^o(t)]^++T_n^o(t) \label{oaque}\\
 Q_n^p(t+1)&=[Q_n^p(t)-R_n^p(t)]^++T_n^p(t) \label{pdque}
\end{align}

All $Q_k$, $Q_n^o$ and $Q_n^p$ have initial values of zero. We define
$
t_n^o\triangleq\lim\limits_{t\rightarrow\infty}\frac{1}{t}\sum_{\tau=0}^{t-1}T_n^o(\tau)$,
$t_n^p\triangleq\lim\limits_{t\rightarrow\infty}\frac{1}{t}\sum_{\tau=0}^{t-1}T_n^p(\tau)
$
as the long-term time-average admission rates of open data and private data respectively. The long-term time-average service rates of  $Q_n^o$ and $Q_n^p$ are also defined as:
 $r_n^o\triangleq\lim\limits_{t\rightarrow\infty}\frac{1}{t}\sum_{\tau=0}^{t-1}R_n^o(\tau)$ and $
r_n^p\triangleq\lim\limits_{t\rightarrow\infty}\frac{1}{t}\sum_{\tau=0}^{t-1}R_n^p(\tau)$. $Q_n^o$ and $Q_n^p$ should be
kept \emph{strongly stable}  in order to ensure the rate requirements of open and private date can be supported by the CR system, which means  $ t_n^o\le r_n^o $ and $t_n^p\le
r_n^p$ hold.

Virtual queues of open { data}, $X_n^o(t)$, and private data $ X_n^p(t) $ are introduced in  (\ref{ovque}) and
(\ref{pvque}) to assist in developing our algorithms, which would guarantee that the actual queues $Q_n^o$ and $Q_n^p$ are
bounded
deterministically  in the worst case.
\begin{align}
  X_n^o(t+1)&=[X_n^o(t)-T_n^o(t) ]^++ \mu_n^o(t) \label{ovque}\\
  X_n^p(t+1)&=[X_n^p(t)-T_n^o(t) ]^++\mu_n^p(t) \label{pvque}
\end{align}
Denote $\mu_n^o$ and $\mu_n^p$ as the virtual admission rates of open data and private data, which are upper bounded  by $D_n^o$ and $D_n^p$ respectively.
Notice that $X_n^o$, $X_n^p$, $\mu_n^o$ and $\mu_n^p$ do not stand for any actual queue and data. They are only generated
by the proposed algorithms. According to queuing  theory, when $X_n^o$ and $X_n^p$  are stable, the  long-term time-average value of
$\mu_n^o$ and $\mu_n^p$  would satisfy:
\begin{align}
\nu_n^o=\lim_{t\rightarrow\infty}\frac{1}{t}\sum_{\tau=0}^{t-1}\mu_n^o(\tau)\le t_n^o\label{xo}\\
\nu_n^p=\lim_{t\rightarrow\infty}\frac{1}{t}\sum_{\tau=0}^{t-1}\mu_n^p(\tau)\le t_n^p\label{xp}
\end{align}

To summarise, as shown in Fig. \ref{sysfig:3} the control space $\mathbf{\chi}$ of the system can be expressed as $\mathbf{\chi}=\{\mathbf{P}^{SU},\mathbf{\Gamma}^{SU},\mathbf{\zeta},\mathbf{T}\}$, which includes admission control $\mathbf{T}=\{T_n^o,T_n^p|\forall n\}$, power control decision $\mathbf{P}^{SU}$, subcarrier assignment $\mathbf{\Gamma}^{SU}$ and security transmission control $\mathbf{\zeta}$.

\subsection{Basic constraints}

\subsubsection{Power consumption constraint}
Let $E\triangleq\sum_{\forall n,\forall m}p_n^m$ as total power consumption of the whole system in one time slot. There exists a
physical peak power limitation $P_{max}$ that $E$ cannot exceed at any time:
\begin{equation}\label{Pmax}
    0\le E\le P_{max}
\end{equation}

The long-term time-average power consumption also has an upper bound $P_{avg}$, which is proposed for energy conservation:
\begin{eqnarray}
  e &\le& P_{avg} \label{pavg}
\end{eqnarray}
where $e=\lim_{t\to\infty}\frac{1}{t}\sum_{\tau=0}^{t-1}\mathbb{E}\{E(\tau)\}$


\subsubsection{Delay-limited model}
The queuing delay is defined as the time a packet waits in a queue until it can be transmitted. Each SU has a
 long-term time-average queuing delay $\rho_n^o$ for its open data transmission. To each SU, it proposes a delay constraint $\rho_n$
as in (\ref{delay}) for its open transmission.
\begin{equation}\label{delay}
    \rho_n^o\le\rho_n
\end{equation}

\section{Problem Formulation}
{ Considering the simplicity and understandability of mathematic analysis, a special case of one single primary link is considered in the following. In the single PU case, the only one PU is indexed with number $0$. In part C of Section V, the general results of multi-PU case are listed for completeness.}
\subsection{Optimization objective and constraints}
 { Following  above descriptions, the objective of this paper is to improve throughput of secondary network while ensuring stability of primary network. So the problem  is formulated as: Maximize the sum weighted admission rates of all SUs and
 stabilize the PU data queue $Q_0$ at the same time. Let $\theta_n$ and $\varphi_n$ for all $n$ be the
 nonnegative weights for private and open data throughput.}  Then the optimal problem can be formulated as:
\begin{eqnarray}
 \textrm{Maximize}&& \sum_{n=1}^N\{\theta_nt_n^p+\varphi_nt_n^o\} \label{obj}\\
   \textrm{Subject to:}&&0\le t_n^p\le\lambda_n^p,\forall n\nonumber\\
  &&0\le t_n^o\le\lambda_n^o ,\forall n\nonumber\\
  &&\mathbf{t}=(t_n^p,t_n^o)\in\mathbf{\Upsilon}\nonumber\\
  && \limsup_{t\rightarrow\infty}\frac{1}{t}\sum_{\tau=0}^{t-1}\mathbb{E}\{Q_0(\tau)\}<\infty\nonumber\\
  &&(\ref{pavg}),(\ref{delay})\nonumber
\end{eqnarray}
where $\mathbf{\Upsilon}$ is the \emph{network capacity region} of secondary links. { Define the service rate vector as $\mathbf{\upsilon}=(r_n^o,r_n^p)$.  The definition of \emph{network capacity region} $\mathbf{\Upsilon}$ is the region of all non-negative service rate vectors $\mathbf{\upsilon}$ for any possible control actions \cite{neely2010stochastic}.} When the CBS takes a kind of control policy under a certain channel condition, the secondary links will have a decided network capacity and the \emph{network capacity region}
is the set of network capacities under all possible control policies and all channel conditions. In the proposed system,
the control policy of CBS should fulfill subcarrier assignment rule (\ref{assign}), peak power constraint (\ref{Pmax}) and
stabilize all queues including actual queues and virtual queues. { So actually, the control policy that can achieve the network capacity region should satisfy  the following constraints:
\begin{equation}
\begin{cases}
&(\ref{assign}),(\ref{Pmax})\\
&\limsup_{t\rightarrow\infty}\frac{1}{t}\sum_{\tau=0}^{t-1}\mathbb{E}\{Q_n^o(\tau)\}<\infty\\
&\limsup_{t\rightarrow\infty}\frac{1}{t}\sum_{\tau=0}^{t-1}\mathbb{E}\{Q_n^p(\tau)\}<\infty\\
&\limsup_{t\rightarrow\infty}\frac{1}{t}\sum_{\tau=0}^{t-1}\mathbb{E}\{X_n^o(\tau)\}<\infty\\
&\limsup_{t\rightarrow\infty}\frac{1}{t}\sum_{\tau=0}^{t-1}\mathbb{E}\{X_n^p(\tau)\}<\infty
\end{cases}\nonumber
\end{equation}}

Theoretically, we can get the optimal solution to (\ref{obj}) if we get the  distribution of the system CSI and external
data arrival rate beforehand. However, this information can not be obtained accurately. In this paper
an online algorithm requiring only current information of queue state and channel state is proposed and will be described
in detail then.
\subsection{Optimality of SU overlay}
Before detailing the control algorithm, it should be specified the conditions that make SU overlay play a positive role
in this cognitive transmission model other than traditional access methods. We focus on presenting a sufficient condition
on overlay for constant channel conditions here, then we will extend it to time-varying situation.

In the case of static network condition, the optimal problem of SUs' weighted throughput is simplified as
\begin{eqnarray}\label{oppr}
\textrm{Maximize:}&\sum_{n=1}^N\{\theta_nr_n^p+\varphi_nr_n^o\}\\
\textrm{Subject to}&r_0^{PU}=\lambda_0\nonumber
\end{eqnarray}
where we only consider the optimal case when $r_0^{PU}=\lambda_0$.  Notice here, the system maximal weighted sum
data rate under full overlay scheme must be greater than or at least no worse than that when SU can only access the
subcarrier which is not occupied by PU. It is easy to understand that full overlay is a more general access
scheme than spectrum overlay which is a special access situation. We can get an intuition that when all subcarriers are assumed
to be accessed by PU, SU data rate would be  positive under full overlay scheme instead of zero under
traditional overlay scheme.  Thus what we want to prove is the sufficient condition of that SUs perform better in
consideration of PU transmission other than accessing the licensed subcarrier roughly. Let $\kappa$ be the fraction of time
that PU is
actively transmitting, thus:
\begin{align}
&r_0^{PU}=\kappa \sum_{m\in\mathbf{\Gamma}_0^{PU}}\log_2(1+\frac{A_0^mP_0^m}{1+a_{0S}^mp_n^m})\\
&r_n^{p}=\{\sum_{m\in\mathbf{\Gamma}_0^{PU}}\{(1-\kappa)[\log_2(1+a_n^mp_n^m)-\log_2(1+b_n^mp_n^m)]^+\nonumber\\
&\quad+\kappa[\log_2(1+\frac{p_{n}^ma_{n}^m}{1+P_0^ma_{nP}^m})-\log_2(1+\frac{p_{n}^mb_{n}^m}{1+P_0^mb_{nP}^m})]^+\}\nonumber\\
&\quad+\sum_{m\in\mathbf{\Gamma}_{SU}}[\log_2(1+a_n^mp_n^m)-\log_2(1+b_n^mp_n^m)]^+\}\zeta_n\\
&r_n^{o}=\sum_{m\in\mathbf{\Gamma}_{SU}}\log_2(1+a_n^mp_n^m)+\sum_{m\in\mathbf{\Gamma}_0^{PU}}[\kappa \log_2(1+\nonumber\\
&\frac{p_{n}^ma_{n}^m}{1+P_0^ma_{nP}^m})+(1-\kappa)
\log_2(1+a_n^mp_n^m)]-r_n^{p}
\end{align}

We have the following lemma:
\begin{Lemma}
In high SINR region, a sufficient condition for full overlay to be optimum in SU $n$ accessing subcarrier $m$ (both
security and open transmission) is:
\begin{equation}
a_{0S}^m\le \min\{C_{nm}^1,C_{nm}^2\},\forall m
\end{equation}
where $C_{nm}^1=b_n^m/[(1+P_0^mb_{nP}^m+b_n^mP_{max})\log_2(1+b_n^mP_{max})],
C_{nm}^2=a_n^m/[(1+P_0^ma_{nP}^m+a_n^mP_{max})\log_2(1+a_n^mP_{max})]$.
\end{Lemma}

We can have an intuitive explanation on  \emph{Lemma 1}, for SU $n$'s accessing subcarrier $m$. If the cross link (from
CBS to primary link) condition is bad enough (worse than weighted CBS-to-SU channel condition $C_{nm}^2$ and weighted
CBS-to-eavesdropper channel condition $C_{nm}^1$), the full overlay scheme would be the optimal scheme when both security
and open transmission happen. The proof of \emph{Lemma 1} can be found in Appendix C.

 It would be obvious to derive the following lemma on sufficient condition of optimality of the whole system overlay.
 Thus we get:
 \begin{Lemma}
 In high SINR region, a sufficient condition for full overlay to be optimum in the whole OFDMA-based CR system is:
\begin{equation}\label{lemma2}
a_{0S}^m\le \min_{n}\{C_{nm}^1,C_{nm}^2\},\forall m
\end{equation}
 \end{Lemma}

Notice that, the sufficient condition does not mean that subcarrier $m\in\boldsymbol{\Gamma}_0^{PU}$ would provide a greater data
rate than $m'\in\boldsymbol{\Gamma}_{SU}$ under the same power allocation scheme. It means that for $m\in
\boldsymbol{\Gamma}_0^{PU}$, full overlay would achieve the optimal result other than any other access policy such as partial
overlay or underlay.
 We assume the sufficient condition of \emph{Lemma 2} is fulfilled in this paper and we proceed considering time-varying
 channels then.
\section{Online Control Algorithm}
It is worth noticing that problem (\ref{obj}) has long-term time-average limitations on power consumption and queuing delay.
Using the technique similar to \cite{neely2010stochastic}, we construct power virtual queue $Y$  and delay virtual queue $
Z_n $ to track the power consumption and queuing delay respectively. These virtual queues do not exist in practice, and
they are just generated by the iterations of (\ref{pque}) and (\ref{dque}):
\begin{align}
  Y(t+1) &= [Y(t)-P_{avg}]^++E(t)\label{pque}\\
  Z_n(t+1) &= [Z_n(t)-\rho_n \mu_n^o]^++Q_n^o(t)\label{dque}
\end{align}

Similar to actual queues, $Y$ and $Z_n$ have initial values of zero. According to \emph{Necessary Condition for Rate
Stability} in \cite{neely2010stochastic}, if  $Y$  is stable, constraint (\ref{pavg}) is satisfied. In addition, if $Z_n$
is stable, $q_n^o=\lim\limits_{t\rightarrow\infty}\frac{1}{t}\sum_{\tau=0}^{t-1}\mathbb{E}\{Q_n^o(\tau)\}\le
\rho_n\nu_n\le \rho_nt_n^o$ holds.  According to Little's Theorem, ${q_n^o}/{t_n^o}=\rho_n^o$,  when $Z_n$ is stable, the
delay constraint (\ref{delay}) would be achieved.
It will be proven that the proposed optimal control algorithm can stabilize these queues in section VI, that is to say the
 long-term time-average constraints are fulfilled.

Using virtual queues $X_n,Z_n$ and $Y$, we decouple problem (\ref{obj}) into two parts: one is \emph{flow control
algorithm} which decides the admission of data, and another is \emph{resource allocation algorithm} in charge of
subcarrier assignment, power allocation and  secure transmission control in every slot. All these control actions aim at
secondary links and happen in CBS. The whole algorithm is named CBS-side online control algorithm (COCA).
\subsection{Flow control algorithm}
When external data arrives at CBS, CBS will decide whether to admit it according to queue lengthes. Let $V$ be a fixed
non-negative control parameter. Let $q_{max}^o\ge \mu_{max}$ and $q_{max}^p\ge D_{max}$ hold. They are actually the
deterministic worst case upper bounds of relative queue length to be proven later. The flow control rules of open data and
private data are obtained by solving (\ref{CTno}) and (\ref{CTnp}) respectively:
  \begin{eqnarray}\label{CTno}
  \textrm{Minimize}&&T_n^o[Q_n^o-q_{max}^o+\mu_{max}]\\
  \textrm{Subject to:}&&0\le T_n^o\le D_n^o\nonumber
\end{eqnarray}
 \begin{eqnarray}\label{CTnp}
  \textrm{Minimize}&& T_n^p[Q_n^p-q_{max}^p+D_{max}]\\
  \textrm{Subject to:}&&0\le T_n^p\le D_n^p\nonumber
 \end{eqnarray}
The corresponding solutions to (\ref{CTno}) and (\ref{CTnp}) are easy to get:
 \begin{equation}\label{tno}
T_{n}^o = \left\{ \begin{array}{ll}
0&\textrm{if} \quad Q_n^o-q_{max}^o+\mu_{max}\ge 0 \\
D_n^o&\textrm{otherwise}
\end{array}\right.
\end{equation}
 \begin{equation}\label{tnp}
T_{n}^p = \left\{ \begin{array}{ll}
0&\textrm{if} \quad Q_n^p-q_{max}^p+D_{max}\ge 0 \\
D_n^p&\textrm{otherwise}
\end{array}\right.
\end{equation}

Here we can have an intuitive explanation on flow control rules. They work like valves. When  any actual data queue
exceeds some threshold, the corresponding valve would turn off and no data would be admitted.

As to virtual variable $\mu_n^o$ and $\mu_n^p$, there are also their respective virtual flow control algorithms
(\ref{muo}) and (\ref{mup}) so as to update virtual queues $X_n^o$ and $X_n^p$ which will play an important role in
resource allocation:
 \begin{eqnarray}\label{muo}
 \textrm{Minimize}&&\mu_n^o[\frac{q_{max}^o-\mu_{max}}{q_{max}^o}X_n^o-\rho_n Z_n-V\varphi_n]\\
 \textrm{Subject to:}&&0\le\mu_n^o\le D_n^o\nonumber\\
 \quad\nonumber\\
\textrm{Minimize}&&\mu_n^p[\frac{q_{max}^p-D_{max}}{q_{max}^p}X_n^p-V\theta_n]\label{mup}\\
\textrm{Subject to:}&&0\le\mu_n^p\le D_n^p\nonumber
\end{eqnarray}

Solutions to (\ref{muo}) and (\ref{mup}) are (\ref{muof}) and (\ref{mupf}) respectively:
 \begin{align}
&\mu_n^o= \left\{ \begin{array}{ll}
0&\textrm{if} \quad(\frac{q_{max}^o-\mu_{max}}{q_{max}^o}X_n^o-\rho_nZ_n-V\varphi_n)\ge 0 \\
D_n^o&\textrm{otherwise}
\end{array}\right.\label{muof}\\
&\mu_n^p= \left\{ \begin{array}{ll}
0&\textrm{if} \quad(\frac{q_{max}^p-D_{max}}{q_{max}^p}X_n^p-V\theta_n)\ge 0 \\
D_n^p&\textrm{otherwise}
\end{array}\right.\label{mupf}
\end{align}

\subsection{Resource allocation algorithm}

The resource allocation policy can be found in solving the following optimization problem.
\begin{align}\label{RA}
&\textrm{Maximize}\quad U(\boldsymbol{P}^{SU},\boldsymbol{\zeta})\\
& \textrm{Subject to:}\quad (\ref{assign}),(\ref{Pmax})\nonumber
\end{align}
where
$U(\boldsymbol{P}^{SU},\boldsymbol{\zeta})=\sum_{n=1}^N(\frac{X_n^oQ_n^o}{q_{max}^o}R_n^o+\frac{X_n^pQ_n^p}{q_{max}^p}R_n^p)+Q_0R_0^{PU}-YE$.

At the beginning of every slot, all $X_n^o,X_n^p,Q_n^o,Q_n^p$ and $Y$ can be regarded as constants because they all have
been decided in the previous slot. $Q_0$ can be estimated by CBS by overhearing PBS feedback. In section VII we propose an
imperfect estimation scheme of $Q_0$ and compare the performances of perfect and imperfect estimations in simulations.
Notice that, the resource allocation is determined at the beginning of every slot and all queues are updated at the end of
every slot.

 Firstly, we can easily decide the vector $\boldsymbol{\zeta}$ maximizing $U$ by assuming that all elements of
 $\boldsymbol{\zeta}$ are continuous variables between 0 and 1 and  in further discussion, we can get a discrete
 implementation of $\zeta_n$.

We take partial derivative in $U(\boldsymbol{P}^{SU},\boldsymbol{\zeta})$ with respect to $\zeta_{n}$:
\begin{align}\label{zderivation}
 &\frac{\partial{U(\boldsymbol{P}^{SU},\boldsymbol{\zeta})}}{\partial{\zeta_n}}=(\frac{X_n^pQ_n^p}{q_{max}^p}-
 \frac{X_n^oQ_n^o}{q_{max}^o})\sum_{m=1}^M\hat{R}_{n}^{mp}
\end{align}
Observing (\ref{zderivation}), $\sum_{m=1}^M\hat{R}_{n}^{mp}$ is no-negative and  $U$ is monotonic in $\zeta_n$, and thus the
optimality condition of secure transmission control is:

\begin{equation}\label{zeta}
\zeta_{n}^* = \left\{ \begin{array}{ll}
1&\textrm{if} \quad (\frac{X_n^pQ_n^p}{q_{max}^p}-\frac{X_n^o Q_n^o}{q_{max}^o})\ge 0 \\
0&\textrm{otherwise}
\end{array}\right.
\end{equation}

Then we use $\boldsymbol{\zeta}^*$  to assign subcarrier and power which is the solution to the following
optimization problem PS,

\begin{align}
\textrm{PS:}\qquad
&\textrm{Maximize}\quad \widetilde{U}(\boldsymbol{P}^{SU})\nonumber\\
& \textrm{Subject to:}\quad (\ref{assign}),(\ref{Pmax})\nonumber
\end{align}
where $\widetilde{U}(\boldsymbol{P}^{SU})=U(\boldsymbol{P}^{SU},\boldsymbol{\zeta}^*)$. PS is a typical Weighted Sum Rate (WSR)
maximization problem,
and it is difficult to find a global optimum since $\widetilde{U}(\boldsymbol{P}^{SU})$ is neither convex nor concave of
$\boldsymbol{P}^{SU}$. Obviously, PS has a typical D.C. structure which can be optimally solved by D.C.
programming\cite{xu2008global}.
In \cite{venturino2009coordinated} there lists a dual decomposition iterative suboptimal algorithm solving this kind of
constrained nonconvex problem instead of D.C. programming. In addition, because of the characteristics of OFDMA networks,
the duality gap is equal to zero even if PS is nonconvex  when the number of subcarriers is close to infinity
\cite{yu2006dual}. So we take a more computationally effective dual method to solve PS and due to space limitation, we
give the key steps here only. 

We define $R_n^{mp}=\zeta_n^*\hat{R}_n^{mp}$ and $R_n^{mo}=C_n^m-R_n^{mp}$. Then the Lagrange function of PS is expressed
as:
\begin{align}\label{lf}
J(\delta,\boldsymbol{P}^{SU})=&\sum_{m=1}^M\{\sum_{n=1}^N[\frac{X_n^o
Q_n^o}{q_{max}^o}R_{n}^{mo}+\frac{X_n^pQ_n^p}{q_{max}^p}R_{n}^{mp}
-Y p_{n}^{m}] \nonumber\\
&+Q_0R_0^m\}+\delta(P_{max}-E)
\end{align}
where $\delta$ is the non-negative Lagrange multiplier for the peak power constraint in problem PS.  The dual problem of
PS is:
$
 \min\limits_{\delta\ge 0} H(\delta)\nonumber
$,
where $H(\delta)=\max\limits_{\boldsymbol{P}^{SU}\ge 0}\{J(\delta,\boldsymbol{P}^{SU})\}$.

When $\delta$ is fixed, we can decide the parameters $\boldsymbol{P}^{SU}$ maximizing the objective of $H(\delta)$. Observing
$H(\delta)$, we find that it can be decoupled into $M$ subproblem as:
\begin{align}
H(\delta)&=\sum_{m=1}^M\max_{\boldsymbol{P}^m_{SU}}J_m(\delta,\boldsymbol{P}^m_{SU})+\delta P_{max}\nonumber\\
&=\sum_{m=1}^M
\max_{\boldsymbol{P}^m_{SU}}\sum_{n=1}^N J_n^m(\delta,p_n^m) +\delta P_{max}
\end{align}
where $\boldsymbol{P}^m_{SU}=\{p_n^m|1\le n\le N\}$, $J_m(\delta,\boldsymbol{P}^m_{SU})=\sum_{m=1}^MJ_n^m(\delta,p_n^m)$,
$J_n^{m}(\delta,p_n^m)=\frac{X_n^o Q_n^o}{q_{max}^o}R_{n}^{mo}+\frac{X_n^pQ_n^p}{q_{max}^p}R_{n}^{mp}
- (Y+\delta ) p_{n}^{m}+Q_0R_0^m(n)$ and
\begin{align}
R_0^m(n)=\left\{\begin{array}{ll}
0&m\in\mathbf{\Gamma}_{SU}\\
R_0^m&m\in\mathbf{\Gamma}_0^{PU}, \varpi_n^m=1
\end{array}\right.
\end{align}

 For $m\in \boldsymbol{\Gamma}_{SU}$, we can get $p^{m*}_{n}$ by taking partial derivative of $J(\delta,\boldsymbol{P}^{SU})$
 with respect to $p_n^m$ and making (\ref{power0}) equal to zero:
\begin{align}\label{power0}
&\textrm{for }m\in \boldsymbol{\Gamma}_{SU}:\nonumber\\
&\frac{\partial(J(\delta,\boldsymbol{P}^{SU}))}{\partial{p_n^m}}= 
\frac{X_n^oQ_n^o}{q_{max}^o}\frac{1}{\ln2}\{\frac{a_n^m}{1+p_n^ma_{n}^m}-\zeta_n[\frac{a_n^m}{1+p_n^ma_{n}^m}\nonumber\\
&-\frac{b_n^m}{1+p_n^mb_n^m}]\}
+\frac{X_n^pQ_n^p}{q_{max}^p}\frac{1}{\ln2}\zeta_n[\frac{a_n^m}{1+p_n^ma_{n}^m}-\frac{b_n^m}{1+p_n^mb_n^m}]\nonumber\\
&-(Y+\delta)
\end{align}

 However, for $m\in \boldsymbol{\Gamma}_{0}^{PU}$,  a  global optimal solution $p^{m*}_{n}$ maximizing $J_n^m$ can be got easily by
 an exhaustive search such as clustering methods or enumerative methods\cite{horst1996global} and it is
 computationally tractable \cite{yu2006dual,cendrillon2006optimal}.

 Substituting (\ref{zeta}) and $p_n^{m*}$ into $J_n^m(\delta,\boldsymbol{P}^{SU})$, the results are denoted as $J_n^{m*}$.
For any subcarrier $m$, it will be assigned to the user who has the biggest $J_{n}^{m*}(\delta,\boldsymbol{P}^{SU})$. Let
$n^*_m$ be the result of subcarrier $m$'s assignment which is given by:
\begin{align}\label{omega}
n^*_m=\arg\max_{n}J_n^m,\forall n \quad\textrm{and}\quad
\varpi^{m*}_{n}=\left\{ \begin{array}{ll}
1&\textrm{if} \quad n=n^*_m \\
0&\textrm{otherwise}
\end{array}\right.
\end{align}

Let $E^*=\sum_{n=1}^N\sum_{m=1}^Mp_{n}^{m*}\varpi^{m*}_{n}$. As to the value of $\delta$, we use subgradient method to
update it as in (\ref{subpower}),
\begin{equation}\label{subpower}
\delta(i+1)=[\delta(i)-\varsigma\triangle\delta(i)]^+
\end{equation}
where $\triangle\delta(i)=P_{max}-E^*(t,i)$.  $\triangle\delta(i)$ is the subgradient of $H(\delta)$  at $\delta$ and
$\varsigma$ is the step size which should be a small positive constant. In addition, index $i$ stands for iteration number.
When the subgradient method converges, the resource allocation is completed.

From the above description, we can find some principles of  resource allocation.

\emph{Remark 1}: In (\ref{zeta}),  both virtual and actual
queues of open as well as private data reflect the gap  between the corresponding user's demand on data rate and the data
rate that the system can provide. Thus, $\frac{X_n^oQ_n^o}{q_{max}^o}$ and $\frac{X_n^pQ_n^p}{q_{max}^p}$ can be regarded as
the transmission urgency of open data and private data. Only when the transmission urgency of private data exceeds open data,
CBS would allocate some resource to transmit private data. Otherwise,  CBS would use the user's entire resource to transmit
open
data due to delay constraint. In PS, it is easy to find that a bigger $Y$ results in less power allocated to every user,
which will reduce the system power consumption. Also we let $Q_0$ to be the weights of $R_0^{PU}$ in PS. It means that if the
transmission pressure of PU is high, CBS will allocate less power in subcarrier set $\boldsymbol{\Gamma}_0^{PU}$ to avoid
causing too much interference on primary link.

{\emph{Remark 2}:
 In the sub-problem of PS, the transmission power of PBS is assumed to be external variables. Even for the worst case that PBS does not control its transmission power actively, the proposed resource allocation algorithm aims to maximize $Q_0R_0^{PU}$ in PS by adjusting the interference from the secondary networks to primary networks. Thus, it can be found that the proposed algorithm actually does not affect the energy consumption of  primary networks too much.  }

 { \subsection{Control algorithm of Multi-PU case}

Flow control algorithm is the same as (\ref{tno}), (\ref{tnp}), (\ref{muof}) and  (\ref{mupf}).

Resource allocation of multi-PU implementation is the solution to problem MPS:
\begin{align}
&\textrm{\boldmath{MPS:} }\nonumber\\
&\textrm{Maximize:}\nonumber\\
&\quad\sum_{n=1}^N\frac{X_n^oQ_n^o}{q_{max}^o}R_n^o+\sum_{n=1}^N\frac{X_n^pQ_n^p}{q_{max}^p}R_n^p+\sum_{k=1}^KQ_kR_k^{PU}-YE\\
& \textrm{Subject to:} \qquad(\ref{assign}),(\ref{Pmax})\nonumber
\end{align}

In next section, the algorithm performance with single PU is analysed. It is easy for  readers to prove that multi-PU implementation ensures primary data queue stability and furthermore enjoys a similar
performance as single PU situation. }
\begin{table}\label{alg}
\renewcommand{\arraystretch}{1.3}
\caption{Algorithm Descriptions}
\centering
\begin{tabular}{|l|}
\hline
{Proposed online control algorithm in timeslot $t$}\\
\hline
\textbf{1)} \underline{Flow control:} \\
\quad Use (\ref{tno}), (\ref{tnp}), (\ref{muof}), (\ref{mupf}) to calculate $T_n^o$, $T_n^p$, $\mu_n^o$ and $\mu_n^p$
\\
\quad respectively.\\
\textbf{2)} \underline{Resource allocation:}\\
\quad \textbf{a)} Set the Lagrange multiplier $\delta=\delta_{ini}$, ($\delta_{ini}$: An initial value of $\delta$).\\
\quad \textbf{b)} For each $(n,m)$\\
\qquad\qquad \romannumeral1) Use (\ref{zeta}) to calculate $\zeta_n^*$.\\
\qquad\qquad \romannumeral2) Use (\ref{power0}) or exhaustive search to find $p_{n}^{m*}$.\\
\qquad\qquad \romannumeral3) Use (\ref{omega}) to calculate $\varpi_{n}^{m*}$.\\
\quad \textbf{c)} Use (\ref{subpower}) to update $\delta$ and calculate $\triangle\delta(i)$.\\
\quad \textbf{d)} If $|\triangle\delta(i)|>\triangle\delta_{c}$, goto \textbf{b)}, else proceed.\\
 \qquad($\triangle\delta_{c}$: converge condition of $\triangle\delta$)\\
\textbf{3)} \underline{Update the queues:}\\
 Use (\ref{oaque}), (\ref{pdque}), (\ref{ovque}), (\ref{pvque}), (\ref{pque}) and (\ref{dque}) to update all queues
 including $Q_n^o,$\\$Q_n^p$, $X_n^o$, $X_n^p$, $Y$, $Z_n$.\\
\hline
\end{tabular}
\end{table}

\section{Algorithm Performance}
 Before the analysis it is necessary to introduce some auxiliary variables. Let
 $\mathbf{t}^*=(t_n^{p,*},t_n^{o,*})$ be the solution to the following problem:

\begin{eqnarray}
\max_{\mathbf{t}:\mathbf{t}\in\mathbf{\Upsilon}}&\sum_{n=1}^N\theta_nt_n^p+\varphi_nt_n^o, \nonumber\\
\textrm{Subject to:}& \quad e\le P_{avg}\nonumber
\end{eqnarray}

And $\mathbf{t}^*(\epsilon)=(t_n^{p,*}(\epsilon),t_n^{o,*}(\epsilon))$ denotes the solution  of:
\begin{eqnarray}
\max\limits_{\mathbf{t}:\mathbf{t}+\epsilon\in\mathbf{\Upsilon}}&\sum_{n=1}^N\theta_nt_n^p+\varphi_nt_n^o \nonumber\\
 \textrm{Subject to:}& \quad e\le P_{avg}\nonumber
\end{eqnarray}

 According to\cite{stolyar2005maximizing}, it is true that:
\begin{equation}\label{ep}
\lim_{\epsilon\rightarrow0}\sum_{n=1}^N\{\theta_nt_n^{p,*}(\epsilon)+\varphi_nt_n^{o,*}(\epsilon)\}=\sum_{n=1}^N\{\theta_nt_n^{p,*}+\varphi_nt_n^{o,*}\}
\end{equation}

The algorithm performance will be listed in \emph{Theorem 1} and \emph{Theorem 2}.

\begin{Theorem}
Employing the proposed algorithm, both actual queues of open data $Q_n^o(t)$ and private data $Q_n^p(t)$ in CBS have
deterministic worst-case bounds:
\begin{eqnarray}
    &&Q_n^o(t)\le q_{max}^o,Q_n^p(t)\le q_{max}^p,\forall t,\forall n\label{optbound1}
\end{eqnarray}

\end{Theorem}

\begin{Theorem}
Given
\begin{align}
   &q_{max}^o>\mu_{max}+\frac{{C_{max}^o}^2+\mu_{max}^2}{2\epsilon},\label{bound1}\\
   &q_{max}^p\ge D_{max}+\frac{{C_{max}^p}^2+D_{max}^2}{2\epsilon},\label{bound2}\\
   &\rho_n>\frac{q_{max}^o}{\nu_n^{o,*}(\epsilon)},\forall n\label{bound3}
\end{align}
where $\epsilon$ is positive and can be chosen arbitrarily close to zero. The proposed algorithm performance is bounded by:
\begin{align}\label{performance}
&\liminf\limits_{t\rightarrow\infty}\frac{1}{t}\sum_{\tau=0}^{t-1}\sum_{n=1}^N\{\theta_nT_n^p(\tau)+\varphi_nT_n^o(\tau)\}\nonumber\\
&\qquad\ge\sum_{n=1}^N\{\varphi_nt_n^{o,*}(\epsilon)+\theta_nt_n^{p,*}(\epsilon)\}-
\frac{B}{V}
\end{align}
where $B$ is a positive constant independent of $V$ and its expression can be found in appendix B.

In addition, the algorithm also ensures that the long-term time-average  sum of PU queue $Q_0$  and virtual queues $X_n^o$, $X_n^p$, $Z_n$,
$Y$ has an upper bound:
\begin{align}\label{avgbound}
   &\limsup\limits_{t\rightarrow\infty}\frac{1}{t}\sum_{\tau=0}^{t-1}\{\sum_{n=1}^N(X_n^o+X_n^p+Z_n)+Y+Q_0\}\nonumber\\
   &\qquad\le
    \frac{B+V\sum\limits_{n=1}^N\{[\theta_nt_n^{p,*}+\varphi_nt_n^{o,*}]\}}{\sigma}
\end{align}
where $o\le\sigma\le\epsilon$. The proof of Theorem 1 is in appendix A. Theorem 2 and the definition of $\sigma$ can be
found in appendix B.
\end{Theorem}

\emph{Remark 3 (Network stability)}: According to the definition of \emph{strongly stability} as shown in
(\ref{stronglystable}), (\ref{optbound1}) and (\ref{avgbound}) indicate the stabilities of all queues in the network
system. As a result, the network system is stabilized and the  long-term time-average constraints of delay and power are satisfied.
Notice here that $Q_0$'s stability is proved means the PU queue stability constraint is fulfilled.  $Q_0$'s stability means that the long-term throughput performance is uninfluenced.  In addition, if PU's arrival rates are within the stability region of PU networks, $Q_0$'s stability can be ensured by the proposed scheduling algorithm for any transmission power of PU base station. Therefore, the transmission power of PU network is not affected in this situation. Furthermore, 
(\ref{optbound1}) states that all the actual queues of open data and private data have deterministic upper bounds, and
this characteristic means that the CBS can accommodate the random arrival packets with finite buffer.

\emph{Remark 4 (Optimal throughput performance)}: (\ref{performance}) states a lower-bound on the weighted throughput that
our algorithm can achieve. Since $B$ is a constant independent of $V$, our algorithm would achieve a weighted throughput
arbitrarily close to $\sum_{n=1}^N\{\varphi_nt_n^{o,*}(\epsilon)+\theta_nt_n^{p,*}(\epsilon)\}$ for some $\epsilon\ge
0$.
Furthermore, given any $\epsilon\ge 0$, we can get a better algorithm performance by choosing a larger $V$ without
improving the  buffer sizes. In addition, as it is shown in (\ref{ep}), when $\epsilon$ tends to zero, our algorithm would
achieve a weighted throughput arbitrarily close to $\sum_{n=1}^N\{\varphi_nt_n^{o,*}+\theta_nt_n^{p,*}\}$ with a
tradeoff
in queue length bounds and long-term time-average delay constraints as shown in (\ref{bound1})-(\ref{bound3}). Thus we can see that
with some certain finite buffer sizes, the proposed algorithm can provide arbitrarily-close-to-optimal performance by choosing
$V$, and $V$'s influence on queue length is shifted from actual queues to virtual queues.

\section{Implementation with Imperfect Estimation}
CBS needs the information of queue length from primary networks to decide the resource allocation among SUs.
\cite{georgiadis2006resource} considers a situation that queue length information is shared among all the nodes, but in
CR environment it is impossible to know the non-cooperative PU's queue information precisely. Compared with
getting perfect information about $Q_k$, it is more realistic to know the time-average packet arrival rate of PUs.
Considering this,
%
%
%
%
%
in this section, we propose an imperfect estimation of $Q_k$ by CBS. And the performance of this estimation will be showed
in simulation section. If the PU $k$ is busy, the estimated queue length in CBS is:
\begin{equation}
\hat{Q}_k(t+1)=[\hat{Q}_k(t)-R_k^{PU}(t)]^++(\lambda_k+\iota)
\end{equation}
where $\iota$ is an over-estimated slack variable to promise primary link stability. CBS can get the precise information
when PU is idle by listening to primary link ACK to find that no power is used to transmit PU $k$'s data packets. In this
situation, $\hat{Q}_k=Q_k=0$ perfectly holds.

As to the control algorithm, we use $\hat{Q}_k$ to substitute $Q_k$ in resource allocation algorithm. For simplicity, we
name this implementation COCA-E (CBS-side online control algorithm with estimated PU queue).

\section{Simulation}
In this section, we firstly simulate COCA performance in an examplary CR system with a single primary link and secondary
network consisting of one CBS, eight SUs and 64 subcarriers.  All weights of open data and private data
are set to be 0.8 and 1 respectively. The main algorithm parameters of secondary network are set as: $P_{avg}=0.8W$,
$P_{max}=1W$, $\rho_n=60, \forall
n$, $q_{max}^o=200$, $q_{max}^p=1000$, $\mu_{max}=50$, $D_{max}=20$  and
$\lambda_n^o=n*0.1*D_{max},\lambda_n^p=n*0.1*\mu_{max}, \textrm{for }n\in\{1,2,\cdots,8\}$.
The  long-term time-average  arrival rate of PU $\lambda_0$ is set to be $140$ and $D^{PU}_{max}=200$. We simulate the multipath channel
of primary and secondary networks as Rayleigh fading channels and the shadowing effect variances are 10 dB. The cross-link
channels between PBS to SUs and CBS to PU are simulated as long-scale fading.
All parameters in the following parts are set the same as these mentioned here, except for other specification.

\begin{figure}
\centering
\includegraphics[width=2.5in, angle=270]{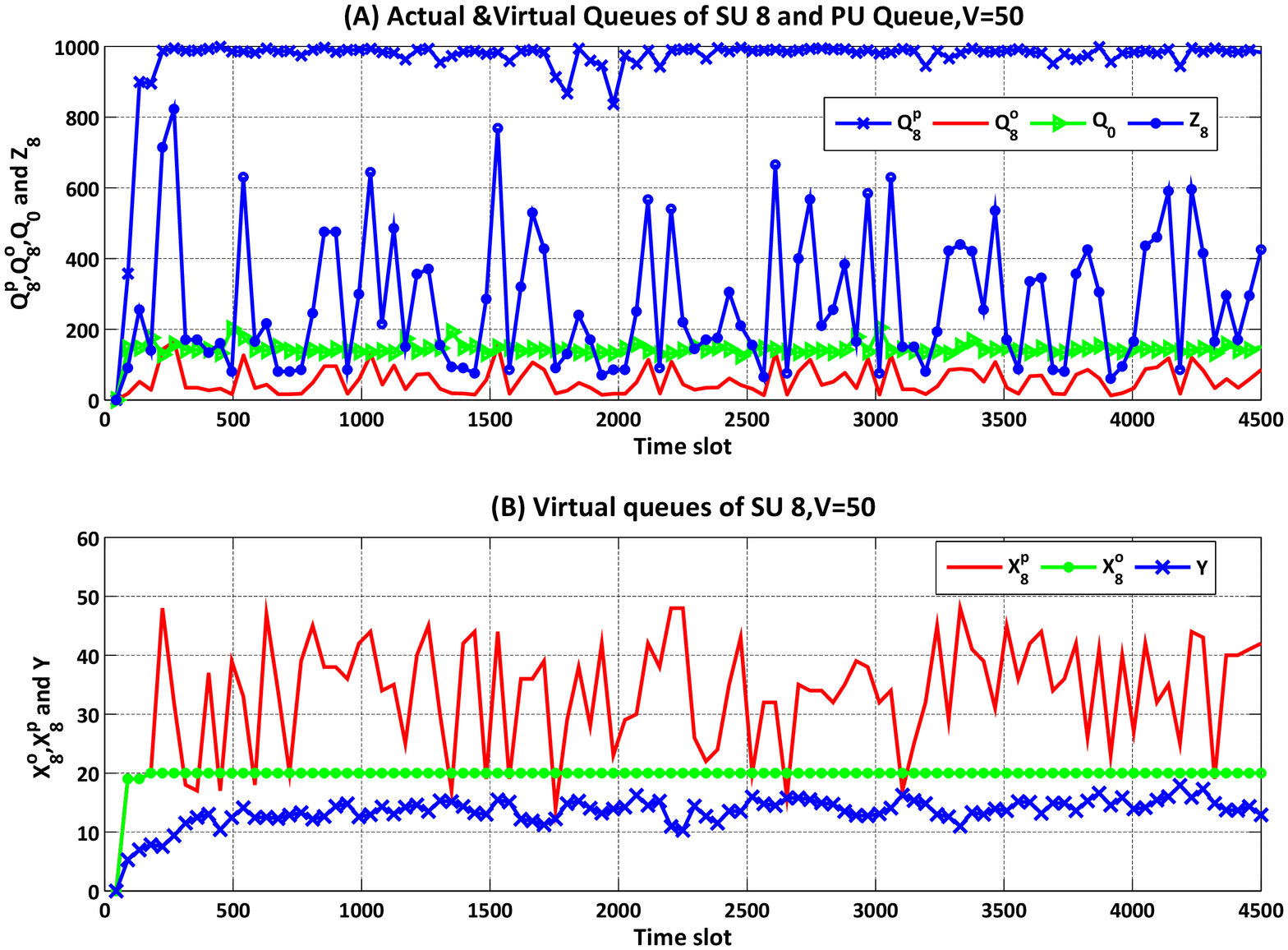}
\caption{Queue evolutions over 4500 slots}
\label{queue_updates}
\end{figure}

\begin{figure}
\centering
\includegraphics[width=2.5in, angle=270]{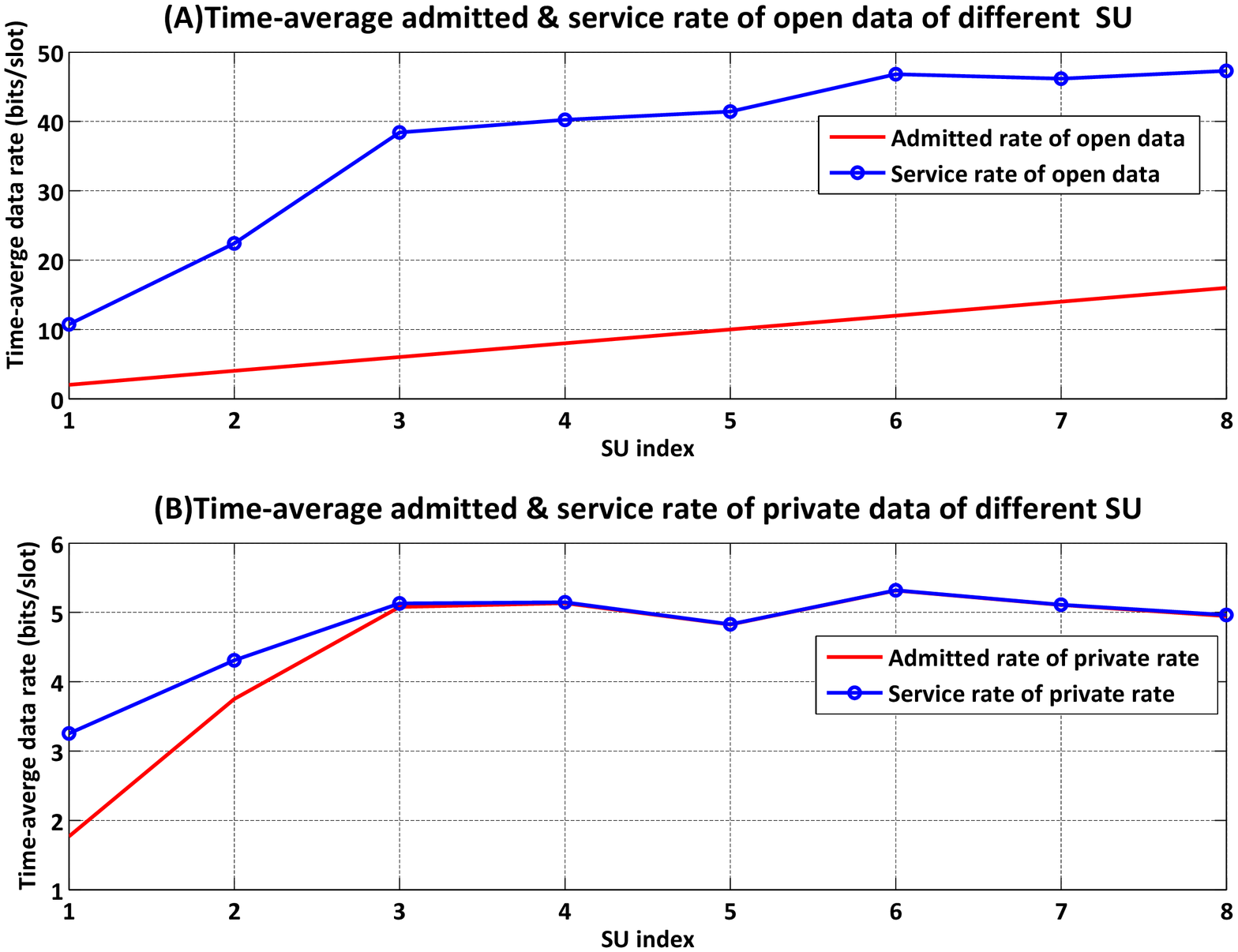}
\caption{Long-term time-average admitted and service rate of all SUs}
\label{ad-ser}
\end{figure}

In Fig. \ref{queue_updates} and Fig. \ref{ad-ser}, we set average value of $a_{0S}^m$, $\overline{a_{0S}}=0.35$,
and average value of $a_{nP}^m$, $\overline{a_{nP}}=21$, $V=50$ and we show both primary and secondary networks'
queue evolution over 4500 slots. Because all SUs' data queues ($Q_n^o,Q_n^p$) and virtual queues ($X_n^o,X_n^p,Z_n$)
enjoy  similar trends,  we take SU 8 as an example.  Fig. \ref{queue_updates} (A) shows the dynamics of SU 8's data queues
$Q_8^o$, $Q_8^p$, virtual delay queue $Z_8$ and PU queue $Q_0$. It is observed that both actual data queues are strictly
lower than their own deterministic worst case upper bound, which verifies \emph{Theorem 1}.  That $Q_0$ is stable
in Fig. \ref{queue_updates} (A) illustrates that our algorithm can ensure PU queues stability from simulation aspect. Besides, in Fig.
\ref{queue_updates}(B), we can also see that virtual queues $X_8^o$, $X_8^p$ and $Y$ are bounded. So  Fig. \ref{queue_updates}
shows that all queues are bounded, which means that the network system is stabilized and the long-term time-average constraints of
delay and power are satisfied.

Fig. \ref{ad-ser} directly shows eight SUs'  long-term time-average admitted rates and service rates of open data and private data,
respectively.  Notice that, every user's admitted rate is smaller than service rate and this promises the
stabilities of actual data queues.

\begin{figure}
\centering
\includegraphics[width=2.5in, angle=270]{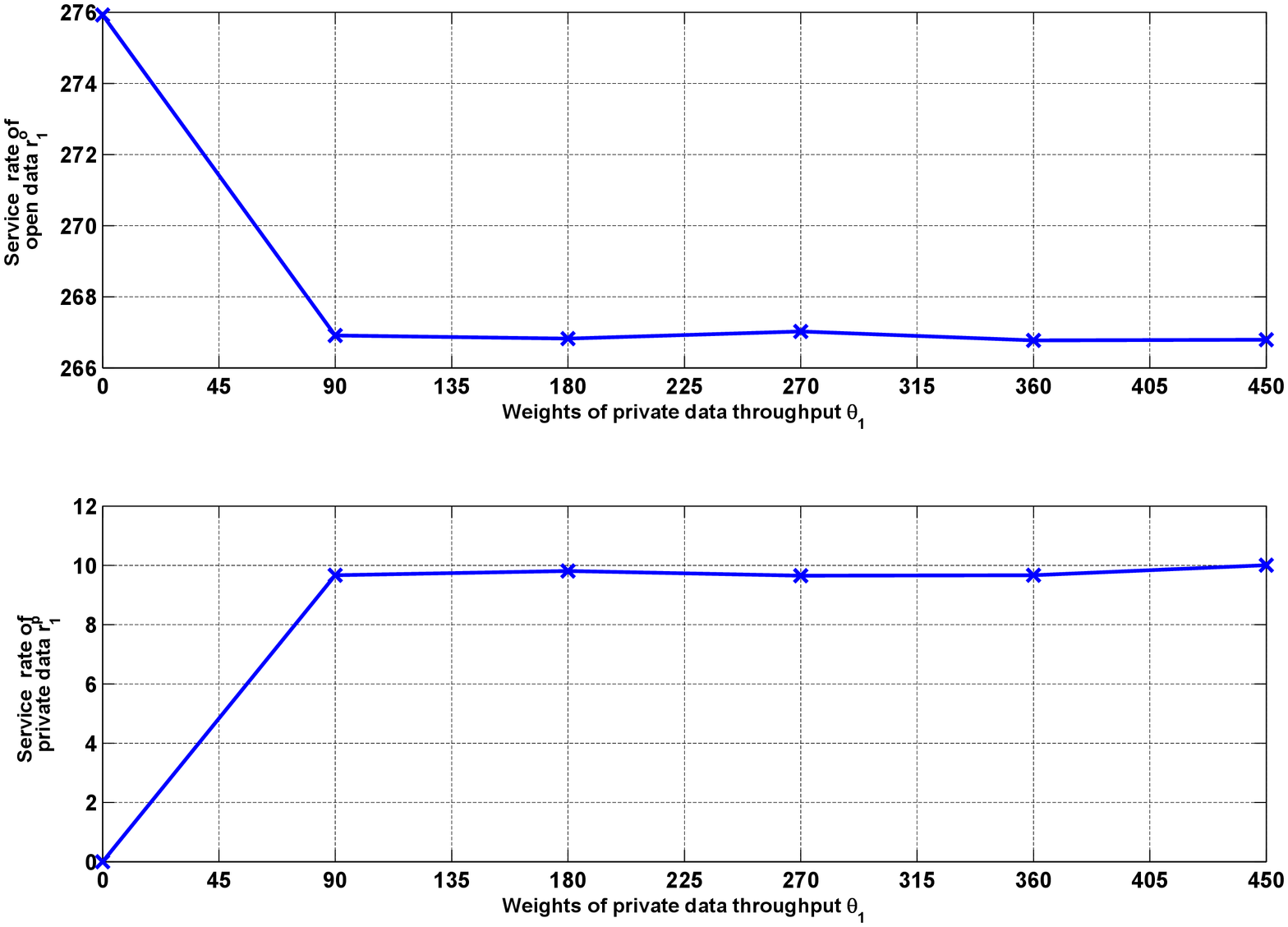}
\caption{ The time average service rate ($r_1^p$ and $r_1^o$) versus the weights of private data $\theta_1$ }
\label{theta-r}
\end{figure}

{ Fig. \ref{theta-r} shows the relationship between the weighting parameters and  long-term time-average service rates. To show the effects more clearly, we consider the scenario consisting of only one SU and one PU with fixed $\varphi_1=450$  and  variational $\theta_1\in \{0,90.180,270,360,450\}$. The  long-term time-average arrival rate of SU is set as: $\lambda_1^p=8$ and $\lambda_1^o=260$. The control parameter $V$ is set to be $380$.  Each value in Fig. \ref{theta-r} is obtained by averaging the converged results of 5000 times. Fig. \ref{theta-r}  shows with the increase of $\theta_1$ the long-term time-average  service rate  of private data increases while the one of open data decreases, which illustrates the effect of throughput weights on long-term time-average  service rates. }

\begin{figure}
\centering
\includegraphics[width=2.5in, angle=270]{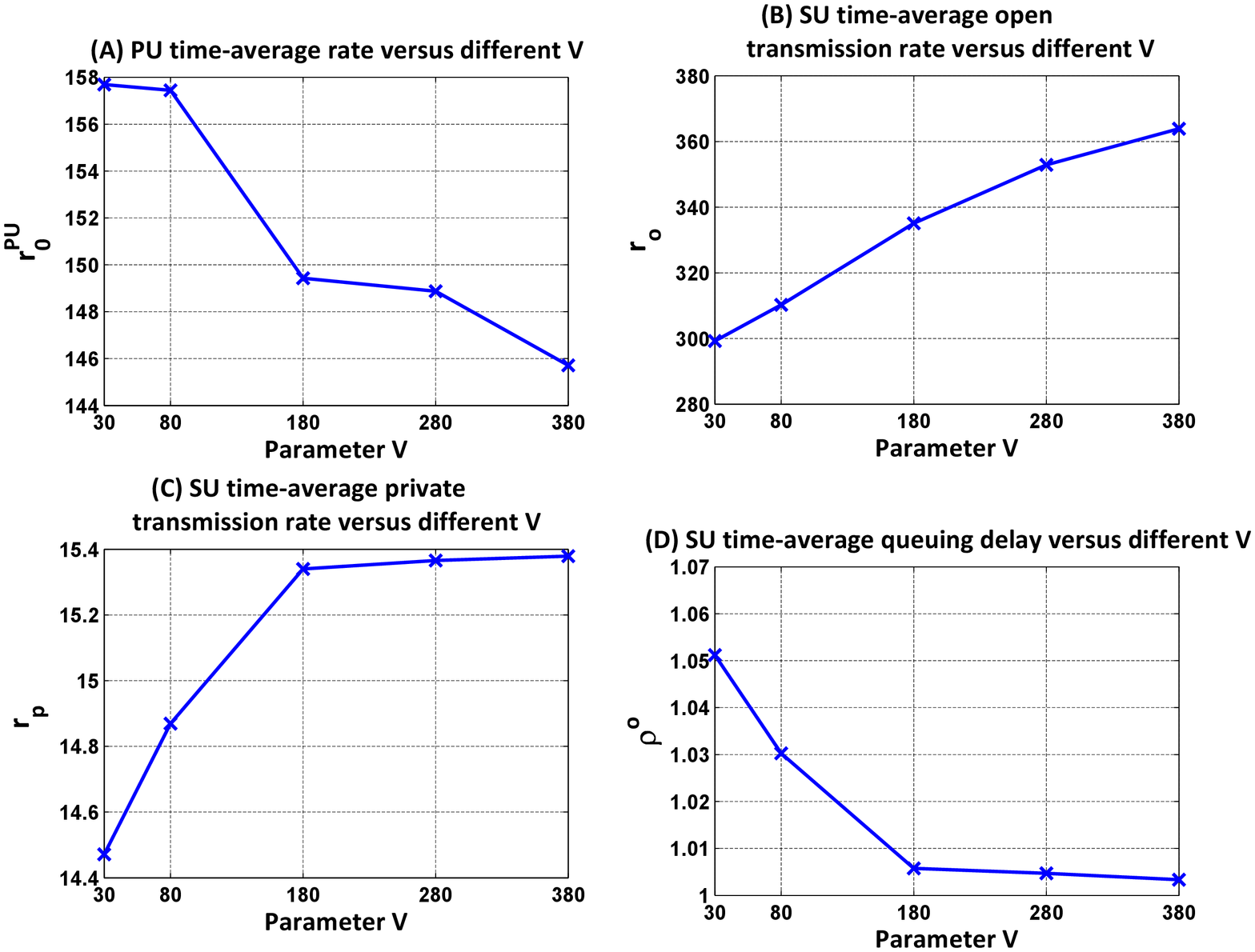}
\caption{ COCA performances (long-term time-average PU rate, SU rates and SU queuing delay) versus control parameter $V$. }
\label{V-perform}
\end{figure}

 Fig. \ref{V-perform}  demonstrates the relationship between different long-term time-average network performance versus control
 parameter $V$. In order to compare PU and SU performance, the similar scenario including one PU and one SU is also considered here. The average data arrival rates of SU  are set as: $\lambda_1^o=250$ and $\lambda_1^p=10$.  In general, the bigger  $V$ results in the
 higher SU open and private transmission rates as  Fig. \ref{V-perform} (B) and  Fig. \ref{V-perform} (C) respectively show.
 Fig. \ref{V-perform} (A) demonstrates PU transmission rate decreases as $V$ increases.  Notice here, although $r_0^{PU}$ decreases, even when $V=380$, $r_0^{PU}$
 approximates $146$ and is greater than $\lambda_0=140$, which preserves PU queue stability. Fig. \ref{V-perform} (D) shows
 the
 queuing delay performance also improves as $V$ increases.

\begin{figure}
\centering
\includegraphics[width=2.5in, angle=270]{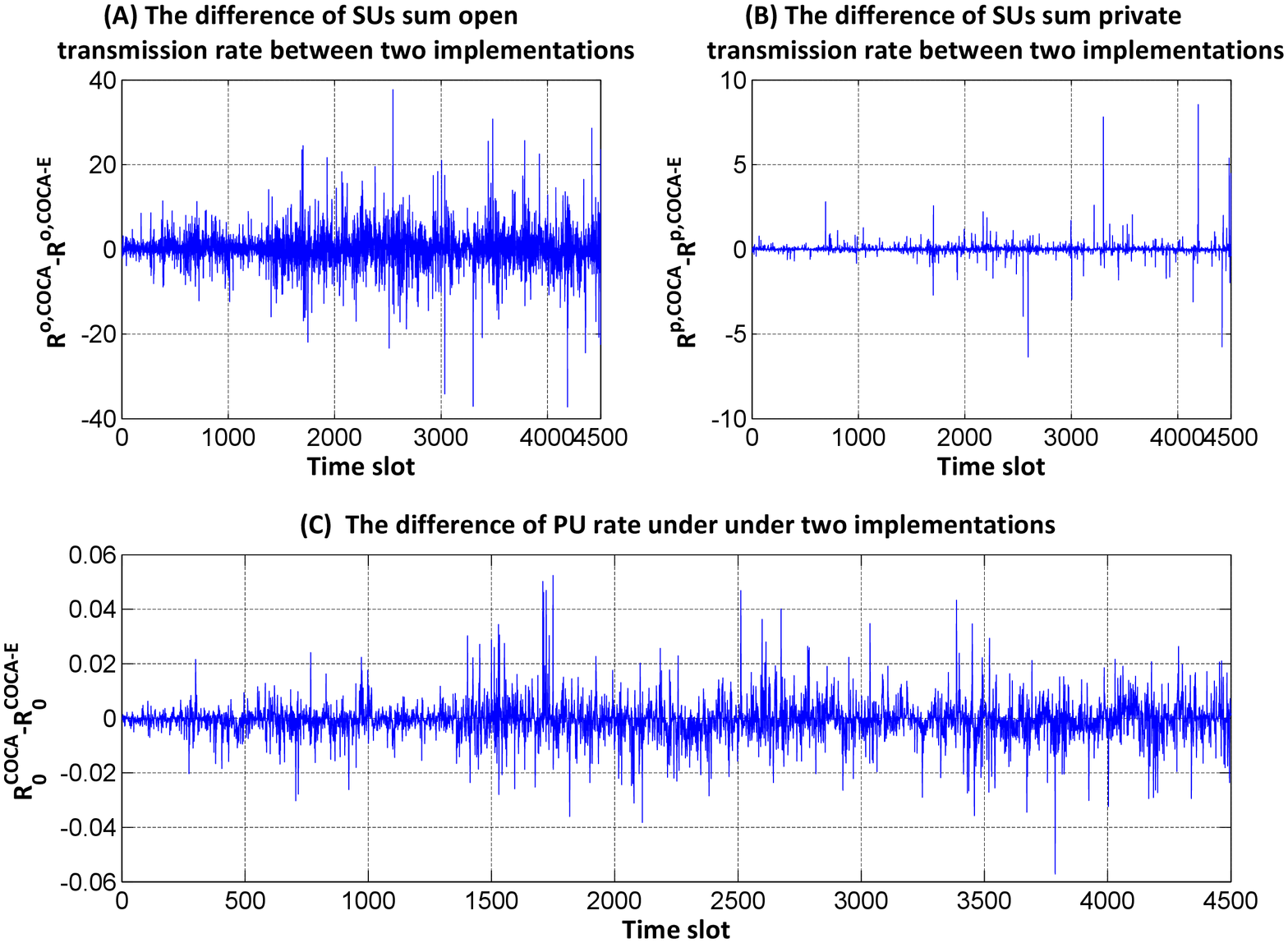}
\caption{ The rate difference of COCA and COCA-E implementation during 4500 slots. }
\label{rate-diff}
\end{figure}

\begin{figure}
\centering
\includegraphics[width=2.5in, angle=270]{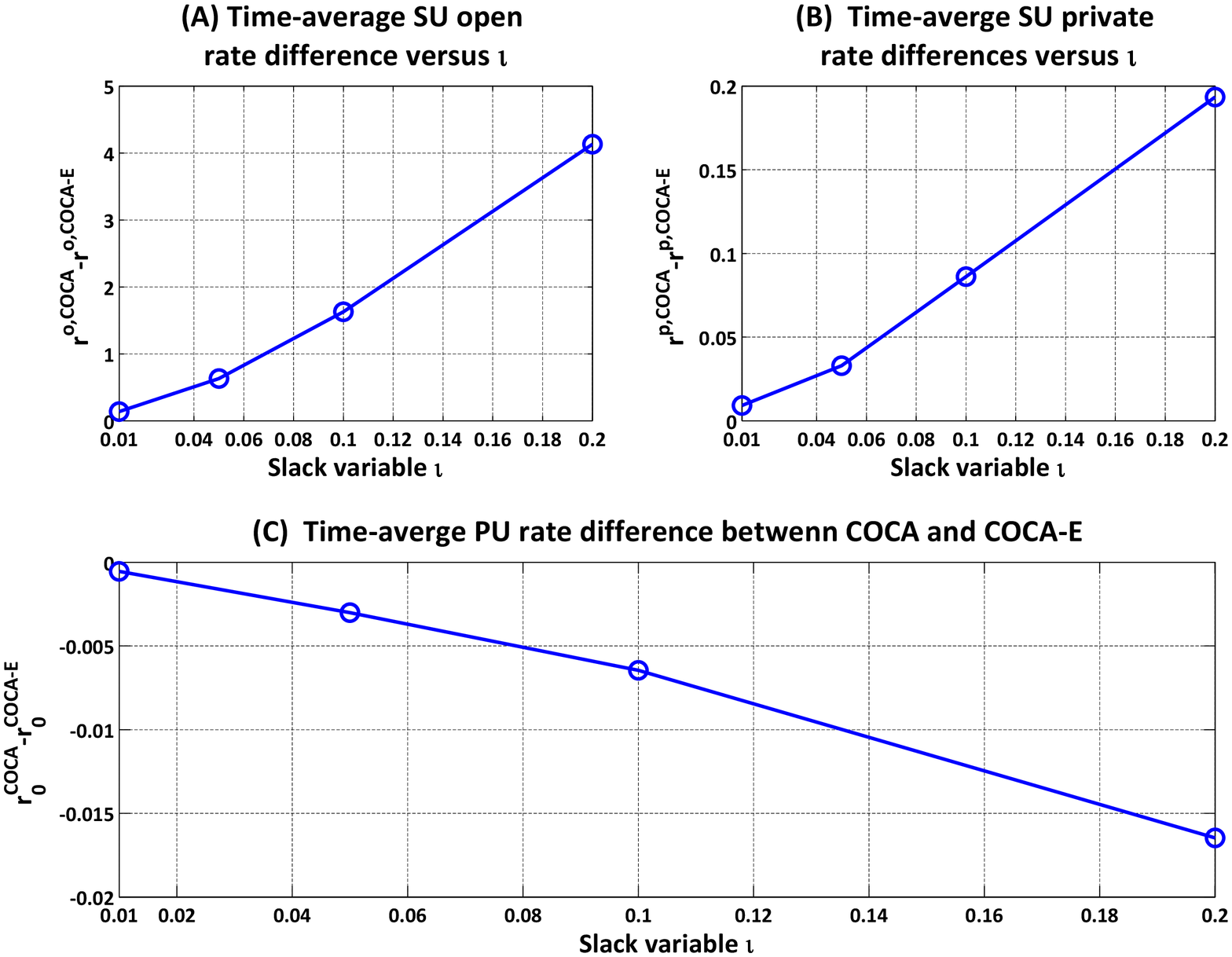}
\caption{ The long-term time-average rate differences of COCA and COCA-E implementation versus different over-estimated slack
variable $\iota$ . }
\label{epsilon-rate}
\end{figure}

The implementation of COCA-E with imperfect estimated $Q_0$ is simulated.  We set the over-estimated slack
variable $\iota$ to be 0.01. We show the differences of the sum service rate of SUs and $R_0^{PU}$ between COCA and COCA-E in Fig.
\ref{rate-diff} (A), Fig. \ref{rate-diff} (B) and Fig. \ref{rate-diff} (C), respectively. We can see that all the
differences are around zero, and SU sum rate is more effected than $R_0^{PU}$ by the imperfect estimation of PU queue information. More
directly, the influence of $\iota$ on  the long-term time-average rate difference between COCA and COCA-E is simulated in Fig.
\ref{epsilon-rate}, where each record is an averaged result of 1000 converged results.  Fig. \ref{epsilon-rate} (C) shows that
$r_0^{COCA}-r_0^{COCA-E}$ becomes more negative as $\iota$ increases,
which means that the rate decline of PU caused by SU transmissions decreases as $\iota$ increases. More directly, if we want
to make sure PU transmission is less influenced, we should choose a larger $\iota$.   While a larger
$\iota$ inevitably  makes SUs' transmission rates decrease as  Fig. \ref{epsilon-rate} (A) and Fig.
\ref{epsilon-rate} (B) show.

\section{Conclusions}
In this paper, we propose a cross-layer scheduling and dynamic spectrum access algorithm for maximizing the long-term average throughput  of open and private information in an OFDMA-based CR network. We derive the sufficient condition to guarantee that full overlay is optimal in this system. The proposed algorithm can provide a flexible scheduling implementation of open and private information while ensuring the stability of primary networks as well as performance requirements in CR systems with finite buffer size. Furthermore, the proposed algorithm is proved to be close to optimality with current network states in time-varying environments.


%

\appendices
\section{Proof of Theorem 1}
Supposing there exists a slot $t$ satisfying  $Q_n^o(t)\le q_{max}^o$, it is obviously true for all queues initialized to
zero. We prove that for $t+1$ the same holds. Obviously, there exists two cases. Firstly, we suppose $Q_n^o(t)\le
q_{max}^o-\mu_{max}$ and we can easily get $Q_n^o(t+1)\le q_{max}^o$. Else, if $Q_n^o(t)>q_{max}^o-\mu_{max}$, then
according to (\ref{tno}), $T_n^o(t)=0$. Then \[Q_n^o(t+1)\le Q_n^o(t)\le q_{max}^o.\]

 The  proof of $Q_n^p\le q_{max}^p$ is similar and omitted here.

\section{Proof of Theorem 2}
Let $\mathbf{Q}=\{Q_0,Q_n^o,Q_n^p,X_n^o,X_n^p,Y,Z_n\}$. We define Lyapunov function $L(\mathbf{Q})$ as:
\begin{align}\label{lyapunov function}
L(\mathbf{Q})=&\frac{1}{2}\{\sum_{n=1}^N[\frac{q_{max}^o-\mu_{max}}{q_{max}^o}{X_n^o}^2+Z_n^2+\frac{1}{q_{max}^o}{Q_n^o}^2(t)X_n^o+\nonumber\\
&\frac{q_{max}^p-D_{max}}{q_{max}^p}{X_n^p}^2+\frac{1}{q_{max}^p}{Q_n^p}^2X_n^p]+Y^2+Q_0^2\}
\end{align}

 According to \cite{neely2010stochastic}, $\triangle L(\mathbf{Q})$ is defined as the conditional Lyapunov drift for slot
 $t$:
 \begin{equation}
\triangle L(\mathbf{Q})\triangleq\mathbb{E}\{ L(\mathbf{Q}(t+1))-L(\mathbf{Q}(t))|\mathbf{Q}(t)\}
 \end{equation}
 According to $(|x-y|+z)^2\le
x^2+y^2+z^2-2x(y-z)$, we can get the results below:

\begin{align}\label{drift1}
    &\frac{q_{max}^o-\mu_{max}}{q_{max}^o}[{X_n^o}^2(t+1)-{X_n^o}^2(t)]\le\nonumber\\
   & \quad\frac{q_{max}^o-\mu_{max}}{q_{max}^o}\{2{\mu_{max}^2}-2X_n^o(t)[T_n^o(t)-\mu_n^o(t)]\}
\end{align}
\begin{align}\label{drift2}
    &\frac{[{Q_n^{o^2}}(t+1)X_n^o(t+1)-{Q_n^{o^2}}(t)X_n^o(t)]}{q_{max}^o}\le q_{max}^o\mu_{max}+\nonumber\\
   &\qquad\frac{(\mu_{max}^2+{C_{max}^{o^2}})-2Q_n^o(t)[R_n^o(t)-T_n^o(t)]}{q_{max}^o}X_n^o(t)
\end{align}

The queues of private data have similar inequalities above. Furthermore, we can derive that:
\begin{align}
&\triangle L(\mathbf{Q})-V\sum_{n=1}^N\mathbb{E}\{\theta_n\mu_n^p+\varphi_n\mu_n^o|\mathbf{Q}\}\le B-Q_0\mathbb{E}\{R_0^{PU}-\nonumber\\
&D_0^{PU}|\mathbf{Q}\}-Y\mathbb{E}\{P_{avg}-E|\mathbf{Q}\}+\sum_{n=1}^N\{\frac{X_n^o({C_{max}^{o^2}}+\mu_{max}^2)}{2q_{max}^o}\notag\\
&+\frac{X_n^p({C_{max}^{p^2}}+D_{max}^2)}{2q_{max}^p}-\frac{X_n^pQ_n^p}{q_{max}^p}\mathbb{E}\{R_n^p-T_n^p|\mathbf{Q}\}-\notag\\
&\frac{X_n^oQ_n^o}{q_{max}^o}\mathbb{E}\{R_n^o-T_n^o|\mathbf{Q}\}-(1-\frac{D_{max}}{q_{max}^p})X_n^p\mathbb{E}\{T_n^p-\mu_n^p|\mathbf{Q}\}-\notag\\
&(1-\frac{\mu_{max}}{q_{max}^o})X_n^o\mathbb{E}\{T_n^o-\mu_n^o|\boldsymbol{Q}\}-Z_n\mathbb{E}\{\rho_n\mu_n^o-Q_n^o|\mathbf{Q}\}\notag\\
&-V\mathbb{E}\{\theta_n\mu_n^p+\varphi_n\mu_n^o|\mathbf{Q}\}\}\label{drift}
\end{align}
where $B=\frac{1}{2}({D^{PU^2}_{max}}+R_{0max}^2+P_{max}^2+P_{avg}^2)+
N[\frac{1}{2}q_{max}^o\mu_{max}+(1-\frac{\mu_{max}}{q_{max}^o})\mu_{max}^2+(1-\frac{D_{max}}{q_{max}^p})D_{max}^2+\frac{1}{2}q_{max}^pD_{max}]
+\frac{1}{2}\sum\limits_{n=1}^N(\rho_n^2\mu_{max}^2+{q_{max}^{o^2}})$
and $C_{max}^p=\max\limits_{n}\{R_{n}^p\},C_{max}^o=\max\limits_{n}\{R_{n}^o\},R_{0max}=\max\{R_0^{PU}\}$. Here we can find
that our algorithm minimizes the right hand side (RHS) of (\ref{drift}).

In order to prove \emph{Theorem 2}, we introduce \emph{Lemma 3}.

\begin{Lemma}
For any feasible rate vector $\mathbf{t}\in\mathbf{\Upsilon}$, there exists a $\mathbf{a}$-only policy $SR$ which stabilizes
the network with the data admitted rate vector, $(\mu_{n,SR}^{p}(t),t_{n,SR}^p(t),\mu_{n,SR}^{o}(t),t_{n,SR}^o(t))$, and the service vector, $(R_{n,SR}^{p}(t),R_{n,SR}^{o}(t))$, independent of data queues. For all $t$ and all $n\in\{1,2,...,N\}$, the flow constraints are satisfied:
\begin{eqnarray}
&&\mathbb{E}\{\mu_{n,SR}^o(t)\}=\mathbb{E}\{T_{n,SR}^{o}(t)\}=\mathbb{E}\{R_{n,SR}^{o}\}\nonumber\\
&&\mathbb{E}\{\mu_{n,SR}^p(t)\}=\mathbb{E}\{T_{n,SR}^{p}(t)\}=\mathbb{E}\{R_{n,SR}^{p}\}\nonumber
\end{eqnarray}

\end{Lemma}

Notice that, the stationary randomized policy $SR$ makes decisions only depending on channel condition and independent of
queue backlogs. Furthermore it may not fulfill the delay constraints. Similar proof of $\mathbf{a}$-only policy is given in \cite{georgiadis2006resource} and the proof of \emph{Lemma 3} is omitted here.

We can control the admitted rate of $\mathbf{t}$ ranging from $\mathbf{t}^*(\epsilon)$ to
$\mathbf{t}^*(\epsilon)+\epsilon$ arbitrarily and resulting in that both $\mathbf{t}^*(\epsilon)$ and $\mathbf{t}^*(\epsilon)+\epsilon$ are
within $\mathbf{\Upsilon}$.
It is assumed that the sufficient condition of full overlay optimum (\ref{lemma2}) is satisfied in our system,  so according to
\emph{Lemma 2}, full overlay can achieve the optimal result. Besides, according to \emph{Lemma 3}, it is true that there
exist two different $\mathbf{a}$-only policies $SR_1$ and $SR_2$ which satisfy:

\begin{align}
&\mathbb{E}\{T_{n,SR_1}^{o}\}=\mathbb{E}\{R_{n,SR_1}^{o}\}=\mathbb{E}\{\mu_{n,SR_1}^o\}=t_n^{o,*}(\epsilon)\label{sr31}\\
&\mathbb{E}\{T_{n,SR_1}^{p}\}=\mathbb{E}\{R_{n,SR_1}^{p}\}=\mathbb{E}\{\mu_{n,SR_1}^{p}\}=t_n^{p,*}(\epsilon)\label{sr32}\\
&\mathbb{E}\{T_{n,SR_2}^{o}\}=\mathbb{E}\{R_{n,SR_2}^{o}\}=\mathbb{E}\{\mu_{n,SR_2}^o\}=t_n^{o,*}(\epsilon)+\epsilon\label{sr33}\\
&\mathbb{E}\{T_{n,SR_2}^{p}\}=\mathbb{E}\{R_{n,SR_2}^{p}\}=\mathbb{E}\{\mu_{n,SR_2}^{p}\}=t_n^{p,*}(\epsilon)+\epsilon\label{sr34}
\end{align}

In addition, for policy $SR_1$ and $SR_2$, it is easy to prove that:
\begin{eqnarray}
&&\mathbb{E}\{R_{0,SR_1}^{PU}\}\ge \lambda_0+\epsilon\label{sr2p1}\\
&&\mathbb{E}\{E_{SR_2}\}\le P_{avg}-\epsilon\label{sr2p2}
\end{eqnarray}

Our algorithm minimizes RHS of (\ref{drift})  among all possible policies including $SR$ policy, thus we can get :
\begin{align}\label{iesr}
&\triangle L(\mathbf{Q})-V\sum_{n=1}^N\mathbb{E}\{\theta_n \mu_n^p+\varphi_n\mu_n^o\}\le B+\notag\\
    &Y\{\mathbb{E}\{E_{SR_2}\}-P_{avg}\}-Q_0\{\mathbb{E}\{R_{0,SR_1}^{PU}\}-\lambda_0\}+\notag\\
    &\sum_{n=1}^N\{Z_nQ_n^o+\frac{{C_{max}^{o^2}}+\mu_{max}^2}{2q_{max}^o}X_n^o+\frac{{C_{max}^{p^2}}+D_{max}^2}{2q_{max}^p}X_n^p+\notag\\
    &[\frac{q_{max}^o-\mu_{max}}{q_{max}^o}X_n^o-Z_n\rho_n-V\varphi_n]X_n^o\mathbb{E}\{\mu^o_{n,SR_1}\}+\notag\\
    &\mathbb{E}\{T^{o}_{n,SR_2}\}\frac{X_n^o}{q_{max}^o}[Q_n^o+\mu_{max}-q_{max}^o]-\frac{X^o_nQ^o_n}{q_{max}^o}\mathbb{E}\{R^o_{n,SR_2}\}+\notag\\
    &\mathbb{E}\{T^{p}_{n,SR_2}\}\frac{X_n^p}{q_{max}^p}[Q_n^p-q_{max}^p+D_{max}]-\frac{X_n^pQ_n^p}{q_{max}^p}\mathbb{E}\{R_{n,SR_2}^p\}+\notag\\
    &\mathbb{E}\{\mu^p_{n,SR_1}\}[\frac{q_{max}^p-D_{max}}{q_{max}^p}X_n^p-V\theta_n]\big\}
\end{align}

After substituting (\ref{sr31})-(\ref{sr34}) , (\ref{sr2p1}) and (\ref{sr2p2}) into the RHS of (\ref{iesr}) and
transforming  it,  we can derive that:
\begin{align}\label{drsr}
\triangle &L(Q)-V\sum_{n=1}^N\mathbb{E}\{\theta_n \mu_n^p+\varphi_n\mu_n^o\}\le B-\epsilon
(Y+Q_0)-\nonumber\\
&\sum\{t_n^{o,*}(\epsilon)\rho-q_{max}\}Z-V\sum_{n=1}^N\{\varphi_nt_n^{o,*}(\epsilon)+\theta_nt_n^{p,*}(\epsilon)\}-\notag\\
&\sum_{n=1}^N\frac{X_n^o}{q_{max}^o}\{\epsilon(q_{max}^o-\mu_{max})-\frac{{C_{max}^{o^2}}+\mu_{max}^2}{2}\}-\nonumber\\
&\sum_{n=1}^N\frac{X_n^p}{q_{max}^p}\{\epsilon(q_{max}^p-D_{max})-\frac{{C_{max}^{p^2}}+D_{max}^2}{2}\}
\end{align}
So when (\ref{bound1})-(\ref{bound3}) hold, we can find  $\epsilon_1>0$ that
$\epsilon_1\le\frac{\epsilon(q_{max}^o-\mu_{max})-\frac{{C_{max}^{o^2}}+\mu_{max}^2}{2}}{q_{max}^o}$,
$\epsilon_1\le t_n^{o,*}(\epsilon)\rho_n-q_{max}$ and
$\epsilon_1\le\frac{\epsilon(q_{max}^p-D_{max})-\frac{{C_{max}^{p^2}}+D_{max}^2}{2}}{q_{max}^p}$. Thus:
\begin{align}\label{last}
&\triangle L(Q)-V\sum_{n=1}^N\mathbb{E}\{\theta_n \mu_n^p+\varphi_n\mu_n^o\}\le
B-V\sum_{n=1}^N\{\varphi t_n^{o,*}(\epsilon)+\nonumber\\
&\theta_nt_n^{p,*}(\epsilon)\}-\sigma(\sum_{n=1}^N\{X_n^o+X_n^p+Z_n\}+Y+Q_0)
\end{align}
where $\sigma=\min\{\epsilon,\epsilon_1\}$.

It can be got that when (\ref{bound1}), (\ref{bound2}) and (\ref{bound3}) hold, (\ref{avgbound}) and
\begin{align}\label{nu_upbound}
&\liminf\limits_{t\rightarrow\infty}\frac{1}{t}\sum_{\tau=0}^{t-1}\sum_{n=1}^N\{\theta_n\mu_n^p(\tau)+\varphi_n\mu_n^o(\tau)\}\ge\nonumber\\
&\qquad\sum_{n=1}^N\{\varphi_nt_n^{o,*}(\epsilon)+\theta_nt_n^{p,*}(\epsilon)\}-
\frac{B}{V}
 \end{align}
 are satisfied by applying the theorem of Lyapunov Optimization, \emph{Theorem 4.2} in \cite{neely2010stochastic}, on (\ref{last})
directly.  Furthermore, (\ref{avgbound}) implies that (\ref{xo}) and (\ref{xp}) hold since $X_n^o$ and $X_n^p$ are kept stable.  So after substituting  (\ref{xo}) and (\ref{xp}) into (\ref{nu_upbound}),  (\ref{performance}) holds. Hence the proof of \emph{Theorem 2} is completed.

\section{Proof of Lemma 1}
From the constraint $r_0^{PU}=\lambda_0$, we can represent $\kappa$ as a function of  $p_n^m$ and substitute it  into
(\ref{oppr}). Then, the optimum solution can be found by solving:
\begin{align}\label{fulloverlay}
\max_{p_n^m}\quad&\lambda_0f(\mathbf{P}^{SU})+\sum_{n=1}^N\sum_{m=1}^M\zeta_n\{(\theta_n-\varphi_n)[\log_2(1+a_n^mp_n^m)\nonumber\\
&-\log_2(1+b_n^mp_n^m)]^+
+\varphi_nW \log_2(1+a_n^mp_n^m)\}\\
\textrm{s.t.}\quad&0\le\sum_{\forall n, \forall m}p_n^m\le P_{max}\nonumber
\end{align}
where
$f(\mathbf{P}^{SU})=\sum_{n=1}^N\{(\theta_n-\varphi_n)\zeta_n\{\sum_{m\in\mathbf{\Gamma}_0^{PU}}[\log_2(1+\frac{p_{n}^ma_{n}^m}{1+P_0^ma_{nP}^m})
-\log_2(1+\frac{p_{n}^mb_{n}^m}{1+P_0^mb_{nP}^m})]^+-\sum_{m\in\mathbf{\Gamma}_0^{PU}}[\log_2(1+a_n^mp_n^m)-\log_2(1+b_n^mp_n^m)]^+\}+\varphi_n\sum_{m\in\mathbf{\Gamma}_0^{PU}}[\log_2(1+\frac{a_n^mp_n^m}{1+P_0^ma_{nP}^m})-\log_2(1+a_n^mp_n^m)]^+\}
/\sum_{m\in\mathbf{\Gamma}_0^{PU}}\{\log_2(1+\frac{A_0^mP_0^m}{1+a_{0S}^mp_n^m})\}$ and  we denote the denominator of
$f(p)$ as $\Delta_d$.

It is reasonable to make an approximation of $f(\mathbf{P}^{SU})$ under the assumption that PU is in a high SINR region such
that $\log_2(1+a_{0S}^mp_n^m+A_0^mP_0^m)\approx \log_2(1+A_0^mP_0^m)$.

 One sufficient condition of full overlay scheme achieving the optimal solution of (\ref{fulloverlay}) is that $\kappa=1$
 makes the objective of (\ref{fulloverlay1}) greater than any other $\kappa\ge0$.
\begin{align}\label{fulloverlay1}
\max_{0<p_n^m\le P^*}&
\lambda_0f(\mathbf{P}^{SU})+\sum_{n=1}^N\sum_{m=1}^M\zeta_n\{(\theta_n-\varphi_n)[\log_2(1+a_n^mp_n^m)\nonumber\\
&-\log_2(1+b_n^mp_n^m)]^+
+\varphi_nW \log_2(1+a_n^mp_n^m)\}
\end{align}
where $P^*\le P_{max}$ is any positive upper bound of $p_n^m$.

Thus we need to find a condition where the maximization solution of (\ref{fulloverlay1})  is an increasing function of
$p_n^m,\forall m \in \mathbf{\Gamma}_0^{PU}$. As $p_n^m$ increases, it will force $\kappa$ to increase as well, eventually
reaching $\kappa=1$, which is full overlay.

Firstly, we analyze the situation when $\zeta_n=1$ and $a_n^m>b_n^m$. In this situation, security transmission
happens with
positive private transmission rate and
$[\log(1+p_n^ma_n^n)-\log(1+p_n^mb_n^n)]^+=\log(1+p_n^ma_n^n)-\log(1+p_n^mb_n^n)$. We derive the first
derivative of the objective of (\ref{fulloverlay1}):
\begin{align}\label{old}
(\theta_n-\varphi_n)&(\frac{a_n^m}{1+a_n^mp_n^m}-\frac{b_n^m}{1+b_n^mp_n^m})+\nonumber\\
&\varphi_nW\frac{a_n^m}{1+a_n^mp_n^m}+\lambda_0\frac{\partial f(\mathbf{P}^{SU})}{\partial p_n^m}
\end{align}

Then for any $m\in\mathbf{\Gamma}_0^{PU}$, we derive that:
\begin{align}\label{old1}
&\frac{\partial f(\mathbf{P}^{SU})}{\partial p_n^m}=\frac{a_{0S}^m}{{\Delta_d}^2(1+
a_{0S}^mp_n^m)}
\{\theta_n\{\log_2\frac{(1+b_n^mp_n^m)}{(1+a_n^mp_n^m)}+\nonumber\\
&\qquad[\log_2(1+\frac{p_{n}^ma_{n}^m}{1+P_0^ma_{nP}^m})-\log_2(1+\frac{p_{n}^mb_{n}^m}{1+P_0^mb_{nP}^m})]\}+\nonumber\\
&\qquad\varphi_n[\log_2(1+\frac{p_{n}^ma_{n}^m}{1+P_0^ma_{nP}^m})
-\log_2(1+\frac{p_{n}^mb_{n}^m}{1+P_0^mb_{nP}^m})]\}+\nonumber\\
&\qquad\frac{1}{\Delta_d}\{\varphi_n[\frac{b_n^m}{1+P_0^mb_{nP}^m+b_n^mp_n^m}-\frac{b_n^m}{1+b_n^mp_n^m}]+\nonumber\\
&\qquad\theta_n[(\frac{b_n^m}{1+b_n^mp_n^m}-\frac{a_n^m}{1+a_n^mp_n^m})+(\frac{a_n^m}{1+P_0^ma_{nP}^m+a_n^mp_n^m}-\nonumber\\
&\qquad\frac{b_n^m}{1+P_0^mb_{nP}^m+b_n^mp_n^m})]\}
\end{align}

Substituting (\ref{old1})  into (\ref{old}), we can get the numeration of the first derivative of the objective of (\ref{fulloverlay1}):
\begin{align}\label{ineq}
&\Delta_d[\theta_n(\Delta_d-\lambda_0)(\frac{a_n^m}{1+a_n^mp_n^m}-\frac{b_n^m}{1+b_n^mp_n^m})+\nonumber\\
&\theta_n\lambda_0(\frac{a_n^m}{1+P_0^ma_{nP}^m+a_n^mp_n^m}-\frac{b_n^m}{1+P_0^mb_{nP}^m+b_n^mp_n^m})+\nonumber\\
&\varphi_n(\Delta_d-\lambda_0)\frac{b_n^m}{1+b_n^mp_n^m}+\varphi_n\lambda_0\frac{b_n^m}{1+P_0^mb_{nP}^m+b_n^mp_n^m}]+\nonumber\\
&\varphi_n[\log_2(1+\frac{b_n^mp_n^m}{1+P_0^mb_{nP}^m})-\log_2(1+b_n^mp_n^m)]+\nonumber\\
&\frac{a_{0S}^m}{1+a_{0S}^mp_n^m}\lambda_0\{\theta_n\{[\log_2(1+\frac{p_{n}^ma_{n}^m}{1+P_0^ma_{nP}^m})-\nonumber\\
&\log_2(1+\frac{p_{n}^mb_{n}^m}{1+P_0^mb_{nP}^m})]
\}\}-\log_2\frac{(1+a_n^mp_n^m)}{(1+b_n^mp_n^m)}\nonumber\\
\ge&\Delta_d[\theta_n\lambda_0(\frac{a_n^m}{1+P_0^ma_{nP}^m+a_n^mp_n^m}-\frac{b_n^m}{1+P_0^mb_{nP}^m+b_n^mp_n^m})+\nonumber\\
&\varphi_n\lambda_0\frac{b_n^m}{1+P_0^mb_{nP}^m+b_n^mp_n^m}]-a_{0S}^m\lambda_0\{\theta_n[\log_2(1+a_n^mp_n^m)-\nonumber\\
&\log_2(1+b_n^mp_n^m)]+\varphi_n\log_2(1+b_n^mp_n^m)\}\nonumber\\
\ge&\theta_n\lambda_0(\frac{a_n^m}{1+P_0^ma_{nP}^m+a_n^mp_n^m}-\frac{b_n^m}{1+P_0^mb_{nP}^m+b_n^mp_n^m})+\nonumber\\
&\varphi_n\lambda_0\frac{b_n^m}{1+P_0^mb_{nP}^m+b_n^mp_n^m}-a_{0S}^m\lambda_0\varphi_n\log_2(1+b_n^mp_n^m)-\nonumber\\
&a_{0S}^m\lambda_0\theta_n[\log_2(1+a_n^mp_n^m)-\log_2(1+b_n^mp_n^m)]
\end{align}

The last inequality  in (\ref{ineq}) holds under the assumption that $A_0^m \gg a_{0S}^m$.

For any pair of $(\theta_n,\varphi_n)$, the RHS of the last inequality of (\ref{ineq}) is greater than zero if and only if
\begin{align}
(\frac{a_n^m}{1+P_0^ma_{nP}^m+a_n^mp_n^m}-\frac{b_n^m}{1+P_0^mb_{nP}^m+b_n^mp_n^m})
-&\nonumber\\
a_{0S}^m[\log_2(1+a_n^mp_n^m)-\log_2(1+b_n^mp_n^m)]&\ge0\\
\frac{b_n^m}{1+P_0^mb_{nP}^m+b_n^mp_n^m}-a_{0S}^m\log_2(1+b_n^mp_n^m)&\ge0
\end{align}

Sufficient conditions for full overlay are:

\begin{align}\label{lem-con}
a_{0S}^m\le&\bar{C}_{mn}^3=\frac{\frac{a_n^m}{1+P_0^ma_{nP}^m+a_n^mp_n^m}-\frac{b_n^m}{1+P_0^mb_{nP}^m+b_n^mp_n^m}}{\log_2(1+a_n^mp_n^m)-\log_2(1+b_n^mp_n^m)}\\
a_{0S}^m\le&\bar{C}_{mn}^1=\frac{\frac{b_n^m}{1+P_0^ma_{nP}^m+b_n^mp_n^m}}{\log_2(1+b_n^mp_n^m)}
\end{align}

 So the sufficient condition of full overlay when $\zeta_n=1$ and $a_n^m>b_n^m$is:
\begin{equation}\label{sc1}
a_{0S}^m\le\min\{\bar{C}_{nm}^3, \bar{C}_{nm}^1\}
\end{equation}

And for $\zeta_n=0$ or $a_n^m\le b_n^m$ we can get similar conclusion and omit the process here. The
sufficient condition is:
\begin{eqnarray}
a_{0S}^m\le \bar{C}_{nm}^2=\frac{\frac{a_n^m}{1+P_0^ma_{nP}^m+a_n^mp_n^m}}{\log_2(1+a_n^mp_n^m)}
\end{eqnarray}

It is easy to find that for any $a_{0S}^m\le \min \{\bar{C}_{nm}^1,\bar{C}_{nm}^2\}$, $a_{0S}^m$ is definitely no
greater than $\bar{C}_{nm}^3$. So for any $m$, the sufficient condition for the optimum of user $n$ accessing this
subcarrier $m$ in full overlay mode is $a_{0S}^m\le \min\{\bar{C}_{nm}^1,\bar{C}_{nm}^2\}$

For $P_{max}>p_n^m$, $C_{nm}^1\le \bar{C}_{nm}^1$ and $C_{nm}^2\le \bar{C}_{nm}^2$ hold. Hence the results in \emph{Lemma 1} are proved.

%
%

\ifCLASSOPTIONcaptionsoff
  \newpage
\fi



\bibliographystyle{IEEEtran}
\bibliography{IEEEabrv,zxz}

\begin{thebibliography}{10}
\providecommand{\url}[1]{#1}
\csname url@samestyle\endcsname
\providecommand{\newblock}{\relax}
\providecommand{\bibinfo}[2]{#2}
\providecommand{\BIBentrySTDinterwordspacing}{\spaceskip=0pt\relax}
\providecommand{\BIBentryALTinterwordstretchfactor}{4}
\providecommand{\BIBentryALTinterwordspacing}{\spaceskip=\fontdimen2\font plus
\BIBentryALTinterwordstretchfactor\fontdimen3\font minus
  \fontdimen4\font\relax}
\providecommand{\BIBforeignlanguage}[2]{{%
\expandafter\ifx\csname l@#1\endcsname\relax
\typeout{** WARNING: IEEEtran.bst: No hyphenation pattern has been}%
\typeout{** loaded for the language `#1'. Using the pattern for}%
\typeout{** the default language instead.}%
\else
\language=\csname l@#1\endcsname
\fi
#2}}
\providecommand{\BIBdecl}{\relax}
\BIBdecl

\bibitem{weiss2004spectrum}
T.~A. Weiss and F.~K. Jondral, ``Spectrum pooling: an innovative strategy for
  the enhancement of spectrum efficiency,'' \emph{Communications Magazine,
  IEEE}, vol.~42, no.~3, pp. S8--14, 2004.

\bibitem{huang2011coolest}
X.~Huang, D.~Lu, P.~Li, and Y.~Fang, ``Coolest path: spectrum mobility aware
  routing metrics in cognitive ad hoc networks,'' in \emph{Distributed
  Computing Systems (ICDCS), 2011 31st International Conference on}.\hskip 1em
  plus 0.5em minus 0.4em\relax IEEE, 2011, pp. 182--191.

\bibitem{6464545}
R.~Deng, J.~Chen, X.~Cao, Y.~Zhang, S.~Maharjan, and S.~Gjessing,
  ``Sensing-performance tradeoff in cognitive radio enabled smart grid,''
  \emph{Smart Grid, IEEE Transactions on}, vol.~4, no.~1, pp. 302--310, March
  2013.

\bibitem{6099643}
R.~Deng, J.~Chen, C.~Yuen, P.~Cheng, and Y.~Sun, ``Energy-efficient cooperative
  spectrum sensing by optimal scheduling in sensor-aided cognitive radio
  networks,'' \emph{Vehicular Technology, IEEE Transactions on}, vol.~61,
  no.~2, pp. 716--725, Feb 2012.

\bibitem{lawrey1999multiuser}
E.~Lawrey, ``Multiuser ofdm,'' in \emph{Signal Processing and Its Applications,
  1999. ISSPA'99. Proceedings of the Fifth International Symposium on},
  vol.~2.\hskip 1em plus 0.5em minus 0.4em\relax IEEE, 1999, pp. 761--764.

\bibitem{zhou2011multiuser}
X.~Zhou, G.~Y. Li, and G.~Sun, ``Multiuser spectral precoding for ofdm-based
  cognitive radios,'' in \emph{Global Telecommunications Conference (GLOBECOM
  2011), 2011 IEEE}.\hskip 1em plus 0.5em minus 0.4em\relax IEEE, 2011, pp.
  1--5.

\bibitem{zhou2011low}
------, ``Low-complexity spectrum shaping for ofdm-based cognitive radios,'' in
  \emph{Wireless Communications and Networking Conference (WCNC), 2011
  IEEE}.\hskip 1em plus 0.5em minus 0.4em\relax IEEE, 2011, pp. 1471--1475.

\bibitem{almalfouh2011interference}
S.~M. Almalfouh and G.~L. Stuber, ``Interference-aware radio resource
  allocation in ofdma-based cognitive radio networks,'' \emph{Vehicular
  Technology, IEEE Transactions on}, vol.~60, no.~4, pp. 1699--1713, 2011.

\bibitem{zhang2009resource}
Y.~Zhang and C.~Leung, ``Resource allocation for non-real-time services in
  ofdm-based cognitive radio systems,'' \emph{Communications Letters, IEEE},
  vol.~13, no.~1, pp. 16--18, 2009.

\bibitem{zhang2009cross}
------, ``Cross-layer resource allocation for mixed services in multiuser
  ofdm-based cognitive radio systems,'' \emph{Vehicular Technology, IEEE
  Transactions on}, vol.~58, no.~8, pp. 4605--4619, 2009.

\bibitem{wang2011advances}
B.~Wang and K.~Liu, ``Advances in cognitive radio networks: A survey,''
  \emph{Selected Topics in Signal Processing, IEEE Journal of}, vol.~5, no.~1,
  pp. 5--23, 2011.

\bibitem{zhao2007survey}
Q.~Zhao and B.~M. Sadler, ``A survey of dynamic spectrum access,'' \emph{Signal
  Processing Magazine, IEEE}, vol.~24, no.~3, pp. 79--89, 2007.

\bibitem{huang2005spectrum}
J.~Huang, R.~A. Berry, and M.~L. Honig, ``Spectrum sharing with distributed
  interference compensation,'' in \emph{New Frontiers in Dynamic Spectrum
  Access Networks, 2005. DySPAN 2005. 2005 First IEEE International Symposium
  on}.\hskip 1em plus 0.5em minus 0.4em\relax IEEE, 2005, pp. 88--93.

\bibitem{le2007qos}
L.~Le and E.~Hossain, ``Qos-aware spectrum sharing in cognitive wireless
  networks,'' in \emph{Global Telecommunications Conference, 2007. GLOBECOM'07.
  IEEE}.\hskip 1em plus 0.5em minus 0.4em\relax IEEE, 2007, pp. 3563--3567.

\bibitem{levorato2012cognitive}
M.~Levorato, U.~Mitra, and M.~Zorzi, ``Cognitive interference management in
  retransmission-based wireless networks,'' \emph{Information Theory, IEEE
  Transactions on}, vol.~58, no.~5, pp. 3023--3046, 2012.

\bibitem{huang2010distributed}
S.~Huang, X.~Liu, and Z.~Ding, ``Distributed power control for cognitive user
  access based on primary link control feedback,'' in \emph{INFOCOM, 2010
  Proceedings IEEE}.\hskip 1em plus 0.5em minus 0.4em\relax IEEE, 2010, pp.
  1--9.

\bibitem{georgiadis2006resource}
L.~Georgiadis, M.~J. Neely, and L.~Tassiulas, ``Resource allocation and
  cross-layer control in wireless networks,'' \emph{Foundations and
  Trends{\textregistered} in Networking}, vol.~1, no.~1, pp. 1--144, 2006.

\bibitem{lapiccirella2013distributed}
F.~E. Lapiccirella, X.~Liu, and Z.~Ding, ``Distributed control of multiple
  cognitive radio overlay for primary queue stability,'' \emph{IEEE
  transactions on wireless communications}, vol.~12, no.~1, pp. 112--122, 2013.

\bibitem{jang2003transmit}
J.~Jang and K.~Lee, ``Transmit power adaptation for multiuser ofdm systems,''
  \emph{Selected Areas in Communications, IEEE Journal on}, vol.~21, no.~2, pp.
  171--178, 2003.

\bibitem{kim2005downlink}
S.~W. Kim, B.-S. Kim, and Y.~Fang, ``Downlink and uplink resource allocation in
  ieee 802.11 wireless lans,'' \emph{Vehicular Technology, IEEE Transactions
  on}, vol.~54, no.~1, pp. 320--327, 2005.

\bibitem{seong2006optimal}
K.~Seong, M.~Mohseni, and J.~Cioffi, ``Optimal resource allocation for ofdma
  downlink systems,'' in \emph{Information Theory, 2006 IEEE International
  Symposium on}.\hskip 1em plus 0.5em minus 0.4em\relax IEEE, 2006, pp.
  1394--1398.

\bibitem{zou2010network}
Y.~Zou, T.~Chen, and S.~Li, ``Network-based predictive control of multirate
  systems,'' \emph{IET control theory \& applications}, vol.~4, no.~7, pp.
  1145--1156, 2010.

\bibitem{xingzheng2012flow}
X.~Zhu, J.~Yue, B.~Yang, and X.~Guan, ``Flow rate control and resource
  allocation policy with security requirements in ofdma networks,'' in
  \emph{Intelligent Control and Automation (WCICA), 2012 10th World Congress
  on}.\hskip 1em plus 0.5em minus 0.4em\relax IEEE, 2012, pp. 1020--1025.

\bibitem{shen2005adaptive}
Z.~Shen, J.~Andrews, and B.~Evans, ``Adaptive resource allocation in multiuser
  ofdm systems with proportional rate constraints,'' \emph{Wireless
  Communications, IEEE Transactions on}, vol.~4, no.~6, pp. 2726--2737, 2005.

\bibitem{li2003dynamic}
G.~Li and H.~Liu, ``Dynamic resource allocation with finite buffer constraint
  in broadband ofdma networks,'' in \emph{Wireless Communications and
  Networking, 2003. WCNC 2003. 2003 IEEE}, vol.~2.\hskip 1em plus 0.5em minus
  0.4em\relax IEEE, 2003, pp. 1037--1042.

\bibitem{huang2008multiconstrained}
X.~Huang and Y.~Fang, ``Multiconstrained qos multipath routing in wireless
  sensor networks,'' \emph{Wireless Networks}, vol.~14, no.~4, pp. 465--478,
  2008.

\bibitem{cui2012survey}
Y.~Cui, V.~Lau, R.~Wang, H.~Huang, and S.~Zhang, ``A survey on delay-aware
  resource control for wireless systems¡ªlarge deviation theory, stochastic
  lyapunov drift, and distributed stochastic learning,'' \emph{Information
  Theory, IEEE Transactions on}, vol.~58, no.~3, pp. 1677--1701, 2012.

\bibitem{5572437}
Z.~Yuanyuan, L.~Shaoyuan, and N.~Yugang, ``Networked predictive control of
  constrained linear systems with stability guarantee,'' in \emph{Control
  Conference (CCC), 2010 29th Chinese}, July 2010, pp. 4355--4360.

\bibitem{xue2010delay}
D.~Xue and E.~Ekici, ``Delay-guaranteed cross-layer scheduling in multi-hop
  wireless networks,'' \emph{arXiv preprint arXiv:1009.4954}, 2010.

\bibitem{urgaonkar2009delay}
R.~Urgaonkar and M.~Neely, ``Delay-limited cooperative communication with
  reliability constraints in wireless networks,'' in \emph{INFOCOM 2009,
  IEEE}.\hskip 1em plus 0.5em minus 0.4em\relax IEEE, 2009, pp. 2561--2565.

\bibitem{shannon1949communication}
C.~Shannon, ``Communication theory of secrecy systems,'' \emph{Bell system
  technical journal}, vol.~28, no.~4, pp. 656--715, 1949.

\bibitem{ozarow1985wire}
L.~Ozarow and A.~Wyner, ``Wire-tap channel ii,'' in \emph{Advances in
  Cryptology}.\hskip 1em plus 0.5em minus 0.4em\relax Springer, 1985, pp.
  33--50.

\bibitem{ng2012energy}
D.~W.~K. Ng, E.~S. Lo, and R.~Schober, ``Energy-efficient resource allocation
  for secure ofdma systems,'' \emph{Vehicular Technology, IEEE Transactions
  on}, vol.~61, no.~6, pp. 2572--2585, 2012.

\bibitem{wang2011power}
X.~Wang, M.~Tao, J.~Mo, and Y.~Xu, ``Power and subcarrier allocation for
  physical-layer security in ofdma-based broadband wireless networks,''
  \emph{Information Forensics and Security, IEEE Transactions on}, vol.~6,
  no.~3, pp. 693--702, 2011.

\bibitem{pei2010secure}
Y.~Pei, Y.-C. Liang, L.~Zhang, K.~C. Teh, and K.~H. Li, ``Secure communication
  over miso cognitive radio channels,'' \emph{Wireless Communications, IEEE
  Transactions on}, vol.~9, no.~4, pp. 1494--1502, 2010.

\bibitem{kwon2012secure}
T.~Kwon, V.~W. Wong, and R.~Schober, ``Secure miso cognitive radio system with
  perfect and imperfect csi,'' in \emph{Global Communications Conference
  (GLOBECOM), 2012 IEEE}.\hskip 1em plus 0.5em minus 0.4em\relax IEEE, 2012,
  pp. 1236--1241.

\bibitem{liang2009capacity}
Y.~Liang, A.~Somekh-Baruch, H.~V. Poor, S.~Shamai, and S.~Verd{\'u}, ``Capacity
  of cognitive interference channels with and without secrecy,''
  \emph{Information Theory, IEEE Transactions on}, vol.~55, no.~2, pp.
  604--619, 2009.

\bibitem{ekrem2012capacity}
E.~Ekrem and S.~Ulukus, ``Capacity region of gaussian mimo broadcast channels
  with common and confidential messages,'' \emph{Information Theory, IEEE
  Transactions on}, vol.~58, no.~9, pp. 5669--5680, 2012.

\bibitem{conf/wcnc/ZhuYG13}
X.~Zhu, B.~Yang, and X.~Guan, ``Cross-layer scheduling with secrecy demands in
  delay-aware ofdma network.'' in \emph{Wireless Communications and Networking
  Conference (WCNC), 2013 IEEE.}\hskip 1em plus 0.5em minus 0.4em\relax IEEE,
  2013, pp. 1339--1344.

\bibitem{tse2005fundamentals}
D.~Tse and P.~Viswanath, \emph{Fundamentals of wireless communication}.\hskip
  1em plus 0.5em minus 0.4em\relax Cambridge university press, 2005.

\bibitem{wallace2002method}
M.~Wallace, J.~R. Walton, and A.~Jalali, ``Method and apparatus for measuring
  reporting channel state information in a high efficiency, high performance
  communications system,'' Oct.~29 2002, uS Patent 6,473,467.

\bibitem{suraweera2010capacity}
H.~A. Suraweera, P.~J. Smith, and M.~Shafi, ``Capacity limits and performance
  analysis of cognitive radio with imperfect channel knowledge,''
  \emph{Vehicular Technology, IEEE Transactions on}, vol.~59, no.~4, pp.
  1811--1822, 2010.

\bibitem{mclaughlin2014applications}
\BIBentryALTinterwordspacing
S.~McLaughlin, W.~Harrison, J.~McConnell, and C.~Argon, ``Applications for
  physical-layer security,'' Jun.~19 2014, uS Patent App. 13/962,777. [Online].
  Available: \url{https://www.google.com/patents/US20140171856}
\BIBentrySTDinterwordspacing

\bibitem{mclaughlin2013secure}
S.~W. McLaughlin, D.~Klinc, B.-J. Kwak, and D.~S. Kwon, ``Secure communication
  using error correction codes,'' Jul.~9 2013, uS Patent 8,484,545.

\bibitem{argon2013pre}
C.~Argon, ``Pre-processor for physical layer security.''

\bibitem{emre2011control}
C.~Koksal, O.~Ercetin, and Y.~Sarikaya, ``Control of wireless networks with
  secrecy,'' in \emph{Signals, Systems and Computers (ASILOMAR), 2010
  Conference Record of the Forty Fourth Asilomar Conference on}.\hskip 1em plus
  0.5em minus 0.4em\relax IEEE, 2010, pp. 47--51.

\bibitem{neely2010stochastic}
M.~Neely, ``Stochastic network optimization with application to communication
  and queueing systems,'' \emph{Synthesis Lectures on Communication Networks},
  vol.~3, no.~1, pp. 1--211, 2010.

\bibitem{xu2008global}
Y.~Xu, T.~Le-Ngoc, and S.~Panigrahi, ``Global concave minimization for optimal
  spectrum balancing in multi-user dsl networks,'' \emph{Signal Processing,
  IEEE Transactions on}, vol.~56, no.~7, pp. 2875--2885, 2008.

\bibitem{venturino2009coordinated}
L.~Venturino, N.~Prasad, and X.~Wang, ``Coordinated scheduling and power
  allocation in downlink multicell ofdma networks,'' \emph{Vehicular
  Technology, IEEE Transactions on}, vol.~58, no.~6, pp. 2835--2848, 2009.

\bibitem{yu2006dual}
W.~Yu and R.~Lui, ``Dual methods for nonconvex spectrum optimization of
  multicarrier systems,'' \emph{Communications, IEEE Transactions on}, vol.~54,
  no.~7, pp. 1310--1322, 2006.

\bibitem{horst1996global}
R.~Horst and H.~Tuy, \emph{Global optimization: Deterministic
  approaches}.\hskip 1em plus 0.5em minus 0.4em\relax Springer, 1996.

\bibitem{cendrillon2006optimal}
R.~Cendrillon, W.~Yu, M.~Moonen, J.~Verlinden, and T.~Bostoen, ``Optimal
  multiuser spectrum balancing for digital subscriber lines,''
  \emph{Communications, IEEE Transactions on}, vol.~54, no.~5, pp. 922--933,
  2006.

\bibitem{stolyar2005maximizing}
A.~Stolyar, ``Maximizing queueing network utility subject to stability: Greedy
  primal-dual algorithm,'' \emph{Queueing Systems}, vol.~50, no.~4, pp.
  401--457.

\end{thebibliography}
\end{document}